\documentclass[
reprint,
aps,
pre,
amsmath,
amssymb,
groupedaddress,
longbibliography,
floatfix
]{revtex4-2}

\usepackage{graphicx}
\graphicspath{{figures/}}
\usepackage{bm}
\usepackage{booktabs}
\usepackage{mathtools}
\usepackage{physics}
\usepackage{hyperref}
\hypersetup{hidelinks}

\begin{document}

\title{Self-consistent field theory of semiflexible nematics: Density–nematic coupling, anisotropic elasticity, and defect core sizes}
\author{Longyu Qing}
\email{qing0006@umn.edu}
\affiliation{School of Physics and Astronomy, University of Minnesota, Minneapolis, Minnesota 55455, USA}

\author{Jorge Vi\~nals}
\affiliation{School of Physics and Astronomy, University of Minnesota, Minneapolis, Minnesota 55455, USA}

\date{\today}

\begin{abstract}
Landau--de Gennes theory provides a successful continuum description of nematic order and topological defects, but connecting its phenomenological coefficients to microscopic molecular properties remains challenging in lyotropic liquid crystals. They often exhibit strong elastic anisotropy and, as is the case in lyotropic chromonics, defect cores are much larger than characteristic molecular length scales. A self consistent field theory for fixed length wormlike chains is introduced to describe the effects of chain flexibility and of density--nematic order coupling on equilibrium uniform and defected states. Results show that while the coupling between density and nematic director is weak, a stronger density--degree of order coupling softens collective nematic fluctuations, controls the limit of stability of the nematic phase, and modifies the magnitudes and anisotropy of the correlation lengths. Near a $+1/2$ disclination, density deviations relative to the uniform state are shown to be large, and modify the extent of the core. Further, the director profile is anisotropic. The defect core size is found to scale approximately linearly with the bulk correlation length, a trend also observed in an augmented Landau--de Gennes model. These results provide microscopic insight on how chain density and flexibility couple to nematic order, and point to an augmented Landau--de Gennes model for lyotropic liquid crystals.
\end{abstract}
\maketitle

\section{Introduction}
\label{sec:intro}

Landau--de Gennes (LdG) theory is a successful phenomenological description of nematic order, nematic elasticity, and topological defects \cite{re:shri96,re:dunmur98,selinger2016introduction}. The theory does not by itself specify how molecular structure or additional microscopic degrees of freedom enter its phenomenological coefficients. Such microscopic underpinnings become particularly difficult to describe in the case of lyotropic liquid crystals, not just theoretically but also experimentally \cite{re:zhou17b,re:varytimiadou24}. Lyotropics encompass a broad range of structurally complex systems including polymers, micellar and chromonic aggregates in solution, nanoparticle colloids, and biological filament systems~\cite{re:zhang18,re:varytimiadou24}. Their response can depend strongly on molecular or aggregate flexibility, and on coupling to additional microscopic degrees of freedom. In compressible lyotropics, nematic order can couple directly to concentration, while in self-assembled chromonic and micellar systems, aggregate scission and recombination give rise to length polydispersity. One prominent manifestation of this microscopic complexity is the large elastic anisotropy generically observed in lyotropic nematics \cite{re:zhou12,re:jeong14,zhou2017fine,re:dietrich20,kim2013morphogenesis}. Another is the defect structure of lyotropic chromonic liquid crystals, where disclination cores can be large enough compared with the characteristic aggregate length scale that they can be observed directly by optical means \cite{zhou2017fine,kim2013morphogenesis}. Such observations raise the question of how molecular flexibility, density redistribution, and additional self-assembly degrees of freedom determine both bulk elastic properties and strongly nonuniform structures such as defects. A central challenge for lyotropic nematics is therefore not only to relate phenomenological coefficients to microscopic structure, but also to determine which additional degrees of freedom, such as density variations or length polydispersity, must be explicitly included in any continuum theory. In addition, low order continuum expansions may become insufficient or inaccurate in strongly ordered or strongly nonuniform regions, such as defect cores, where microscopic structure can become important.

Molecular theories of nematic order commonly model interactions either through an orientation dependent excluded volume term (Onsager theory) \cite{onsager1949effects,de1993physics} or through an effective quadrupolar interaction as in Maier--Saupe (MS) theory~\cite{maier1958einfache,de1993physics}. For semiflexible nematogens, a single molecular orientation is no longer sufficient because conformational degrees of freedom associated with bending and connectivity vary along the contour. These conformational statistics are naturally represented by the wormlike chain model~\cite{kratky1949rontgenuntersuchung}. When combined with self consistent field theory (SCFT) \cite{fredrickson2005equilibrium}, wormlike chain statistics can be coupled self consistently to spatially varying density and orientational order fields, making the framework particularly suitable for the study of nonuniform nematic states. Applications include ordering in confined geometries \cite{chen2016theory}, nematic--isotropic interfaces \cite{qing2025self,jiang2010isotropic}, and topological defects \cite{qing2025self}. 

Within the wormlike chain SCFT framework, orientational interactions may be modeled using either an Onsager type excluded volume interaction, or a MS interaction. Although the two have different microscopic interpretations, both promote nematic alignment, while the simpler local form of the MS interaction makes it particularly convenient for theoretical analysis and numerical calculation \cite{jiang2017thermodynamics,spencer2020nematic,qing2025self}. MS based wormlike chain calculations have already been used to show that chain flexibility strongly affects Frank elastic constants and their anisotropy~\cite{ghosh2022semiflexible,re:macpherson22}. 

In this work, we consider a fixed length wormlike chain model with MS orientational interactions, and an orientation independent segment--segment repulsion. This setting provides a controlled model in which the effects of chain flexibility and density--nematic order coupling can be studied before incorporating reversible aggregation, scission, and the resulting length polydispersity \cite{re:taylor91}. An extension of the present framework to a polydisperse system with reversible self assembly will be presented elsewhere~\cite{re:qing26}.

We first derive an effective free energy for uniform states in terms of segment density and nematic order, which provides a description of phase coexistence and the stability of homogeneous states. This analysis highlights the role that conformational entropy loss due to nematic ordering plays in phase coexistence and stability. We then use the SCFT framework to develop a linear response description of both coupled density--nematic degree of order fluctuations and director distortions about a uniform state. The coupled density--nematic order response yields correlation lengths associated with density and nematic order variations, while the director response yields the Frank elastic constants and shows how chain flexibility controls their anisotropy. Finally, we examine whether these bulk quantities, obtained from small perturbations about a uniform nematic, remain relevant near strongly nonuniform topological defects. This connection is tested for $+1/2$ disclinations by comparing correlation lengths with core sizes, and Frank elastic anisotropy with director configurations.

The remainder of this paper is organized as follows: In Sec.~\ref{sec:theory}, we present the SCFT framework for semiflexible wormlike chains and derive the corresponding self-consistent equations. In Sec.~\ref{sec:uniform}, we construct the effective free energy for uniform states and analyze isotropic--nematic coexistence and stability. In Sec.~\ref{sec:lr}, we develop the linear-response theory for director distortions and coupled density--nematic order variations. Section~\ref{sec:results} presents the numerical results for Frank elasticity and density--nematic correlation lengths, and then examines how these bulk response properties are reflected in the structure of $+1/2$ disclinations. Finally, Sec.~\ref{sec:conclusion} summarizes the main conclusions and discusses directions for future work.

\section{SCFT of wormlike chains}
\label{sec:theory}

We briefly summarize the SCFT framework for semiflexible nematic polymers modeled as wormlike chains with MS interactions. This follows earlier theoretical developments \cite{fredrickson2005equilibrium, spencer2020nematic, chen2016theory, jiang2017thermodynamics}, and our previous work~\cite{qing2025self}. In addition to microscopic interactions, appropriate external fields are introduced to allow a linear response analysis.

We consider a system of $n$ wormlike chains of contour length $L$ and persistence length $l_p$ in a volume $V$ at temperature $T$. The $i$-th chain configuration is described by a space curve $\mathbf{r}_i(s)$ with $s \in [0,L]$, and unit tangent vector $\mathbf{u}_i(s) = d\mathbf{r}_i/ds$. The microscopic, orientation resolved, segment density is defined as
\begin{equation}
\hat{\rho}_u(\mathbf{r},\mathbf{u}) = \sum_{i=1}^{n} \int_0^{L} ds \,
\delta(\mathbf{r}-\mathbf{r}_i(s)) \delta(\mathbf{u}-\mathbf{u}_i(s)),
\end{equation}
where the subscript $u$ labels an orientation-resolved density with respect to the unit tangent $\mathbf u$. This density satisfies the normalization condition $\int d\mathbf{r}\, d\mathbf{u}\, \hat{\rho}_u(\mathbf{r},\mathbf{u}) = n L.$ Its orientational integral defines the microscopic scalar segment density, $\hat{\rho}(\mathbf r) \equiv \int d\mathbf u \, \hat{\rho}_u(\mathbf r,\mathbf u)$, which measures the local contour length per unit volume. We reserve the notation $\hat\rho_u$ for the orientation resolved density and $\hat\rho$ for its orientational integral. For polymer melts, $\hat{\rho}(\mathbf r)$ is naturally interpreted as a density field, whereas in lyotropic solutions it is more naturally interpreted as a concentration field, the two quantities differing only by an overall proportionality constant. For a uniform state, the average scalar density becomes spatially uniform, ${\rho}(\mathbf r)\equiv \langle \hat{\rho}(\mathbf r)\rangle = n L /V$. The Hamiltonian of the system without external fields is given by
\begin{equation}
\begin{aligned}
\beta H_0 =& \frac{l_p}{2} \sum_{i=1}^n \int_0^{L} ds 
\left| \frac{d\mathbf{u}_i}{ds} \right|^2 
\\
&+ \frac{1}{2} \int d\mathbf{r}\, d\mathbf{r}' \, d\mathbf{u}\, d\mathbf{u}' \,
\hat{\rho}_u(\mathbf{r},\mathbf{u}) 
V_I(\mathbf{r},\mathbf{r}';\mathbf{u},\mathbf{u}') 
\hat{\rho}_u(\mathbf{r}',\mathbf{u}'),
\end{aligned}
\end{equation}
where $\beta = 1/(k_B T)$, with $k_B$ the Boltzmann constant. The first term represents the bending energy of individual chains, while the second term accounts for pairwise segment interactions. The interaction kernel is $V_I(\mathbf{r}, \mathbf{r}'; \mathbf{u}, \mathbf{u}')=\delta(\mathbf{r}-\mathbf{r}')\left[u_0-u_2 \, P_2(\mathbf{u} \cdot \mathbf{u}')\right],$ where $u_0$ is the strength of an orientation independent segment--segment repulsion, $u_2$ is the strength of the orientation dependent MS alignment interaction, and $P_2(x)=(3x^2-1)/2$ is the second order Legendre polynomial. In lyotropic and solution-based systems, solvent effects are treated implicitly through the orientation-independent repulsive term. In this interpretation, $u_0$ is a solvent-renormalized interaction parameter related to the underlying Flory--Huggins interaction between segments and solvent particles~\cite{spencer2020nematic, kim2007finite}. Its value depends on solution conditions such as solvent quality, salt concentration, and temperature. Recent experiments on nematogen--solvent mixtures further show that solvent quality can qualitatively alter phase behavior, including the presence or absence of isotropic--isotropic demixing and its interplay with nematic ordering~\cite{bailey2026phase}. In the present effective description, good solvents generally correspond to larger effective repulsion and poor solvents to weaker repulsion. 

An external field is introduced via a linear coupling to $\hat{\rho}_u(\mathbf r,\mathbf u)$,
\begin{equation}
\begin{aligned}
\beta H_{\mathrm{ext}} 
&= \sum_{i=1}^n \int_0^{L} ds \, v(\mathbf{r}_i(s), \mathbf{u}_i(s)) \\
&= \int d\mathbf{r} \, d\mathbf{u} \,
v(\mathbf{r}, \mathbf{u}) \hat{\rho}_u(\mathbf{r}, \mathbf{u}).
\end{aligned}
\end{equation}
The total Hamiltonian is then given by $\beta H = \beta H_0 + \beta H_{\mathrm{ext}}$. The partition function of the system can be expressed as
\begin{equation}
Z=\frac{1}{n! v_0^n}\prod_{i=1}^n \int D^* \mathbf{r}_i \exp(-\beta H),
\end{equation}
where $v_0$ is a reference volume introduced to make the partition function dimensionless. The asterisks on the path integrals indicate that the integrals are performed under the constraint $\prod_{s}\delta \left( \mathbf{u}(s) - \frac{d\mathbf{r}(s)}{ds} \right) \delta(|\mathbf{u}(s)|-1)$. These delta function constraints enforce local inextensibility and unit tangent vectors along the chain contour.

To convert the microscopic partition function into a field theory, we introduce an orientation resolved density field $\rho_u(\mathbf r,\mathbf u)$ by inserting the identity $\int\mathcal D\rho_u\,\delta[\rho_u-\hat\rho_u]=1$. The functional delta constraint is represented through a functional Fourier integral by introducing a conjugate field $w(\mathbf r,\mathbf u)$~\cite{fredrickson2005equilibrium},
\begin{equation}
\begin{aligned}
\delta[\rho_u-\hat\rho_u]
\propto \int\mathcal Dw\,\exp\!\bigg\{&
i\int d\mathbf r\,d\mathbf u\,
w(\mathbf r,\mathbf u)\\[-1mm]
&\times\left[\rho_u(\mathbf r,\mathbf u)-\hat\rho_u(\mathbf r,\mathbf u)\right]
\bigg\}.
\end{aligned}
\end{equation}
The partition function can be written, up to an irrelevant normalization constant, as
\begin{equation}
Z \propto \int \mathcal{D}\rho_u \, \mathcal{D}w \,
\exp\left(-\beta F[\rho_u,w]\right).
\end{equation}
Within the mean-field approximation, fluctuations of the fields $\rho_u$ and $w$ about their stationary values are neglected, and the functional integral is evaluated at the saddle point. The saddle-point solution $w(\mathbf{r},\mathbf{u})$ generally lies off the real axis. Following the standard SCFT convention, we deform the contour through the saddle point and rotate $w\to-iw$, after which the saddle-point field used below is real. The corresponding free energy functional is written, up to additive constants and terms linear in $n$ that only shift the reference chemical potential, as
\begin{equation}
\begin{aligned}
\beta F[\rho_u, w] 
=& - \int d\mathbf{r} \, d\mathbf{u} \,
\bigl[w(\mathbf{r}, \mathbf{u}) - v(\mathbf{r}, \mathbf{u})\bigr]
\, \rho_u(\mathbf{r}, \mathbf{u}) \\
& - n \ln Z_1[w] 
+ n \ln\!\left(\frac{n v_0}{V}\right) \\
 + \frac{1}{2} &\int d\mathbf{r} \, d\mathbf{r}' \, d\mathbf{u} \, d\mathbf{u}' \,
\rho_u(\mathbf{r}, \mathbf{u}) 
V_I(\mathbf{r}, \mathbf{r}'; \mathbf{u}, \mathbf{u}')
\rho_u(\mathbf{r}', \mathbf{u}').
\label{eq:f_functional}
\end{aligned}
\end{equation}
Here, $Z_1[w]$ denotes the single-chain partition function, with the normalization chosen such that $Z_1[0]=1$. In the saddle-point approximation, this free energy is evaluated at its stationary configuration. The stationarity condition with respect to $\rho_u(\mathbf r,\mathbf u)$ gives
\begin{equation}
w(\mathbf r,\mathbf u)
=
\int d\mathbf r' \, d\mathbf u'\,
V_I(\mathbf r,\mathbf r';\mathbf u,\mathbf u')\,
\rho_u(\mathbf r',\mathbf u')
+
v(\mathbf r,\mathbf u),
\label{eq:sc_rho}
\end{equation}
while stationarity with respect to $w(\mathbf r,\mathbf u)$ yields
\begin{equation}
\rho_u(\mathbf r,\mathbf u)
=
-n\,
\frac{\delta \ln Z_1[w]}
{\delta w(\mathbf r,\mathbf u)}.
\label{eq:sc_w}
\end{equation}
Equation (\ref{eq:sc_w}) requires evaluating the single chain partition function in the field $w(\mathbf r,\mathbf u)$~\cite{fredrickson2005equilibrium, spencer2020nematic},
\begin{equation}
\rho_u(\mathbf{r}, \mathbf{u})
= \frac{n}{4\pi V Z_1} 
\int_0^{L} ds \,
q(\mathbf{r}, -\mathbf{u}, L-s;[w]) \,
q(\mathbf{r}, \mathbf{u}, s;[w]),
\end{equation}
where the single chain propagator $q(\mathbf{r}, \mathbf{u}, s;[w])$ satisfies the modified diffusion equation (MDE)
\begin{equation}
\frac{\partial q}{\partial s}
= \left( \frac{1}{2l_p} \nabla^2_{\mathbf{u}} 
- \mathbf{u} \cdot \nabla_{\mathbf{r}} 
- w(\mathbf{r}, \mathbf{u}) \right) q,
\label{eq:mde}
\end{equation}
with initial condition $q(\mathbf{r}, \mathbf{u}, 0;[w]) = 1$. The single chain partition function is given by
\begin{equation}
Z_1[w] 
= \frac{1}{4\pi V} \int d\mathbf{r} \, d\mathbf{u} \,
q(\mathbf{r}, \mathbf{u}, L;[w]).
\end{equation}
At equilibrium, the saddle-point fields $\rho_u(\mathbf{r},\mathbf{u})$ and $w(\mathbf{r},\mathbf{u})$ must simultaneously satisfy the self-consistency equations, Eqs.~(\ref{eq:sc_rho}) and (\ref{eq:sc_w}). In practice, these equations are solved iteratively until convergence is achieved.

From the equilibrium distribution $\rho_u(\mathbf{r},\mathbf{u})$, the orientational order parameter tensor can be computed~\cite{fredrickson2005equilibrium},
\begin{equation}
\mathbf{Q}(\mathbf{r}) = \int d\mathbf{u} \, 
\rho_u(\mathbf{r}, \mathbf{u}) 
\left( \mathbf{u} \otimes \mathbf{u} - \frac{1}{3}\mathbf{I} \right),
\label{eq:tensor_order}
\end{equation}
where $\mathbf{I}$ is the identity tensor. $\mathbf{Q}(\mathbf{r}) $ can be further parametrized as
\begin{equation}
\mathbf{Q}(\mathbf{r}) 
= \rho(\mathbf{r}) \left[
S(\mathbf{r}) \left( \mathbf n \otimes \mathbf n - \frac{1}{3}\mathbf{I} \right)
+ P(\mathbf{r}) \left( \mathbf m \otimes \mathbf m - \mathbf l \otimes \mathbf l \right)
\right],
\end{equation}
where $\mathbf n(\mathbf r)$, $\mathbf m(\mathbf r)$, and $\mathbf l(\mathbf r)$ define a local orthonormal triad, with $\mathbf n(\mathbf r)$ denoting the local nematic director. The scalar fields $S(\mathbf r)$ and $P(\mathbf r)$ denote the uniaxial order parameter and the degree of biaxiality, respectively.

\section{The uniform state}
\label{sec:uniform}
Within the SCFT framework, the equilibrium state is determined by the saddle-point equations \eqref{eq:sc_rho} and \eqref{eq:sc_w}, for $\rho_u(\mathbf r,\mathbf u)$ and $w(\mathbf r,\mathbf u)$. Although these equations provide a complete mean-field description, the dependence of the free energy on the density field $\rho_u$ is obscured by the presence of the auxiliary field $w$. To make this dependence more explicit, we adopt a partial saddle-point construction in which $w$ is eliminated while $\rho_u$ is retained. Specifically, for a prescribed $\rho_u$, we determine the stationary field $w^*[\rho_u]$ by inverting Eq.~(\ref{eq:sc_w}) and neglect fluctuations of $w$ about this stationary solution. This defines the density-dependent effective free-energy functional
\begin{equation}
\beta F_{\rm eff}[\rho_u]
\equiv
\beta F\!\left[\rho_u,w^*[\rho_u]\right].
\end{equation}
The theory can further be coarse grained by expressing the free energy in terms of the low-order moments of $\rho_u(\mathbf r,\mathbf u)$, namely $\rho(\mathbf r)$ and $Q_{ij}(\mathbf r)$. These are the natural coarse-grained variables for the present model because the MS interaction contribution in Eq.~\eqref{eq:f_functional} to the free energy is $\frac12\int d\mathbf r(u_0\rho^2-\frac32 u_2 Q_{ij}Q_{ij}),$ where repeated tensor indices imply summation throughout the paper. 

To characterize bulk thermodynamics independently of spatial variations, we restrict the theory to spatially uniform states in this section. This also provides reference states for the linear response analysis developed in the next section. Among spatially uniform states, we restrict attention to the isotropic and uniaxial nematic phases relevant to the isotropic--nematic transition. For these states, the coarse-grained description further yields an effective free energy as a function of $\rho$ and $S$. We use this free energy to determine isotropic--nematic coexistence from the phase-equilibrium conditions and to characterize the limits of stability of the isotropic and nematic branches at their spinodals.

\subsection{Effective free energy and saddle points}
For spatially uniform uniaxial states, the density is then $\rho=nL/V$, and the nematic order tensor takes the form $Q_{ij}=\rho S\left(n_i n_j-\frac13\delta_{ij}\right).$ The relation in Eq.~\eqref{eq:sc_rho} shows that at a self-consistent solution the interaction generated auxiliary field has the same angular structure as the MS kernel, namely isotropic and quadrupolar components. Motivated by this, we restrict the auxiliary field to the same angular dependence,
\begin{equation}
w(\mathbf u)
=
w_0+w_Q
\left[
(\mathbf u\cdot\mathbf n)^2-\frac13
\right],
\end{equation}
where $w_0$ and $w_Q$ are the amplitudes of the isotropic and quadrupolar components of the auxiliary field, respectively. For the convention adopted here, the usual prolate nematic ordering ($S>0$) corresponds to $w_Q<0$, while oblate ordering ($S<0$) corresponds to $w_Q>0$. Projecting Eq.~\eqref{eq:sc_w} onto the quadrupolar component gives
\begin{equation}
S
=
-\frac{3}{2L}
\frac{\partial\ln Z_1}{\partial w_Q},
\end{equation}
which implicitly defines $w_Q^*(S)$.

For a spatially uniform state, the translational gradient term in the MDE, Eq.~\eqref{eq:mde}, vanishes. Factoring out the $w_0$ contribution through $q(\mathbf u,s)=e^{-w_0 s}\tilde q(\mathbf u,s),$ $\tilde q(\mathbf u,s)$ satisfies
\begin{equation}
\frac{\partial \tilde q(\mathbf u,s)}{\partial s}
=\left[\frac{1}{2l_p}\nabla_{\mathbf u}^2
-w_Q\left(
(\mathbf u\cdot\mathbf n)^2-\frac13\right)\right]
\tilde q(\mathbf u,s),
\label{eq:mde_wq}
\end{equation}
with $\tilde q(\mathbf u,0)=1.$ Therefore, the single chain partition function is $Z_1(w_0,w_Q)=e^{-w_0 L}Z_Q(w_Q),$ where $Z_Q$ denotes the $w_Q$-dependent part. The free energy contribution of $e^{-w_0 L}$ cancels exactly with the term involving $w_0^*\rho$ in the first term of Eq.~\eqref{eq:f_functional}. Using the relation $w_Q^*(S)$, $w_Q^*$ can be eliminated in favor of $S$, thereby expressing the effective free energy density in terms of $\rho$ and $S$,
\begin{equation}
\beta f_{\rm eff}(\rho,S)
=
\frac{\rho}{L}\ln\left(\frac{\rho v_0}{L}\right)
+
\frac12u_0\rho^2
+
\frac{\rho}{L}
\left[
g(S; L/l_p)
-
\frac{\alpha}{2}S^2
\right],
\label{eq:eff_f}
\end{equation}
where $\alpha \equiv u_2\rho L=nL^2u_2/V$ can be interpreted as the effective strength of the alignment interaction. Here,
\begin{equation}
g(S;L/l_p)
=
-\frac23\tilde w_Q^*S
-
\ln Z_Q(\tilde w_Q^*; L/l_p),
\label{eq:g}
\end{equation}
with $\tilde w_Q=Lw_Q$. The function $g(S;L/l_p)$ contains the change in conformational entropy per chain associated with nematic ordering, measured relative to the isotropic reference state ($g(S=0)=0$). The effective free energy form shows that $\rho$ and $S$ are coupled in two ways. First, the prefactor $\rho/L=n/V$ indicates that the free energy density is a sum over single chain contributions. Second, $\rho$ modifies the shape of the $S$-dependent free energy landscape through the effective alignment strength $\alpha$.

The saddle-point solutions are obtained by minimizing the effective free energy with respect to $S$ at fixed $\rho$, giving
\begin{equation}
g'(S;L/l_p)=\alpha S,
\label{eq:extrema}
\end{equation}
where the prime denotes differentiation with respect to $S$. From Eq.\eqref{eq:g}, one has $g'(S;L/l_p)=-\frac{2}{3}\tilde w_Q^*(S).$ Combining these relations, the saddle-point solutions satisfy $\alpha S=-\frac{2}{3}\tilde w_Q^*(S)$.

\subsection{Phase coexistence and limit of stability}
The chemical potential and osmotic pressure follow from the effective free energy density in Eq.~\eqref{eq:eff_f} as
\begin{align}
\beta\mu(\rho,S)
&=
L\frac{\partial(\beta f_{\rm eff})}{\partial\rho}
\nonumber\\
&=
\ln\left(\frac{\rho v_0}{L}\right)
+1
+u_0\rho L
+g
-u_2\rho L S^2,
\nonumber\\[2mm]
\beta \Pi (\rho,S)
&=
\rho\frac{\partial(\beta f_{\rm eff})}{\partial\rho}
-\beta f_{\rm eff}
\nonumber\\
&=
\frac{\rho}{L}
+\frac12u_0\rho^2
-\frac12u_2\rho^2S^2.
\label{eq:mu_pressure}
\end{align}
The isotropic-nematic coexistence condition is obtained by requiring the chemical potential and osmotic pressure to be equal in the two uniform phases. Denoting the isotropic and nematic coexistence densities by $\rho_I$ and $\rho_N$, respectively, we have $S=0$ and $\alpha_I=u_2\rho_I L$ in the isotropic phase, while $S=S_N$ and $\alpha_N=u_2\rho_N L$ in the nematic phase. The nematic order parameter $S_N$ is determined by Eq.~\eqref{eq:extrema} and corresponds to the stable solution with $S_N>0$. The coexistence conditions are therefore
\begin{equation}
\begin{aligned}
\ln\!\left(\frac{\alpha_N}{\alpha_I}\right)
+ g(S_N;L/l_p)
+\frac{u_0}{u_2}(\alpha_N-\alpha_I)
-\alpha_NS_N^2
&=0,
\\
\alpha_N-\alpha_I
+\frac12\frac{u_0}{u_2}
(\alpha_N^2-\alpha_I^2)
-\frac12\alpha_N^2S_N^2
&=0.
\end{aligned}
\label{eq:coexistence}
\end{equation}
For a given $L/l_p$, the coexistence conditions depend only on the single parameter $u_0/u_2$. The resulting coexistence values $\alpha_I$ and $\alpha_N$ then yield the coexistence densities through $\rho_{I,N}=\alpha_{I,N}/(u_2L)$.

The spinodal marks the limit of metastability of a uniform state. Define the Hessian matrix
\begin{equation}
A_{ij}
\equiv
\frac{\partial^2(\beta f_{\rm eff})}
{\partial x_i \partial x_j},
\qquad
x_{i,j} \in \{\rho,S\},
\end{equation}
evaluated at a uniform stationary state. The spinodal is reached when the smallest eigenvalue vanishes. Since $A_{\rho\rho}>0$ throughout the physically relevant parameter range, this condition is equivalent to the condition $\det A =0$ for the present two-variable problem. The isotropic spinodal is given by
\begin{equation}
\alpha_{I,\mathrm{sp}} = g''(S=0;L/l_p),
\end{equation}
where $g''$ denotes the second derivative with respect to $S$. The nematic spinodal is determined by
\begin{equation}
\biggl(
\frac{1}{\alpha_{N,\mathrm{sp}}}
+\frac{u_0}{u_2}
-S_N^2
\biggr)
\left[g''(S_N;L/l_p)-\alpha_{N,\mathrm{sp}}\right]
-\alpha_{N,\mathrm{sp}} S_N^2=0.
\end{equation}
The results show that, for fixed $L/l_p$, the isotropic spinodal occurs at a constant value. In contrast, the nematic spinodal varies with the ratio $u_0/u_2$.

The coexistence and spinodal conditions admit an equivalent formulation in terms of the grand potential density $\beta\omega(\mu,S)=\beta\Omega(\mu,S)/V,$ where $\Omega=Vf_{\rm eff}-\mu n$ is the grand potential. At coexistence, the isotropic and nematic minima of $\beta\omega$ have equal values. The spinodal corresponds to the point at which one of these minima loses stability, signaled by a vanishing curvature with respect to $S$ at fixed chemical potential $\mu$. The equivalence to the Hessian-based stability criterion is shown in Appendix~\ref{app:binodals and spinodals}.

\begin{figure*}[!t]
\centering
\includegraphics[width=0.72\textwidth]{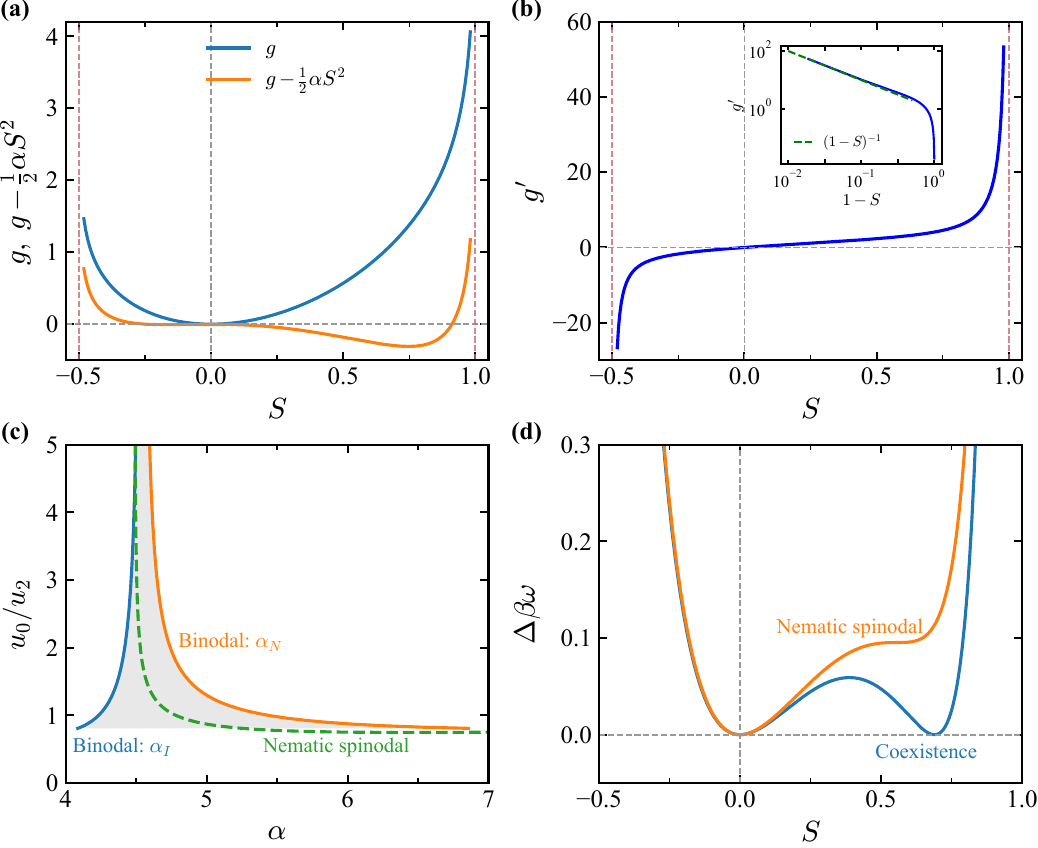}
\caption{
(a) $g(S)$ and $g(S)-\frac12\alpha S^2$ for $\alpha=6$, and (b) $g'(S)$, both in the rigid-chain limit. The outer dashed vertical lines indicate the physical bounds $S=1$ and $S=-1/2$. Inset of (b): log-log plot showing the asymptotic scaling of $g'(S)$ near the perfectly aligned limit. The green dashed line denotes the reference scaling $(1-S)^{-1}$, showing that $g'(S)\sim(1-S)^{-1}$ as $S\to1$. (c) Phase diagram in the rigid-chain limit, showing the isotropic binodal ($\alpha_I$), nematic binodal ($\alpha_N$), and nematic spinodal ($\alpha_{N,\mathrm{sp}}$) as functions of $u_0/u_2$. (d) Grand potential difference $\Delta\beta\omega(S)=\beta\omega(S)-\beta\omega(0)$ for $u_0=u_2=1$ and $L=1$ in the rigid-chain limit. The two curves correspond to chemical potentials chosen at the coexistence point and at the nematic spinodal.}
\label{fig:g_rigid}
\end{figure*}
\subsection{Rigid chain limit}
For finite persistence length $l_p$, Eq.~\eqref{eq:mde_wq} generally requires a numerical solution, for example by expanding the propagator in spherical harmonics and truncating the expansion at finite angular order. As an analytically tractable example of the general free energy construction, we consider the rigid chain limit $l_p\to\infty$. This limit retains the same effective free energy structure and saddle point conditions derived above, while allowing the MDE to be solved analytically. The finite flexibility case differs only through the specific form of $g$, which can be evaluated numerically within the same framework. In the rigid chain limit, the chain propagator is given by
\begin{equation}
\tilde q(\mathbf u,s)
=
\exp\left[
-  w_Q s
\left(
(\mathbf u\cdot\mathbf n)^2-\frac13
\right)
\right].
\end{equation}
Therefore, 
\begin{equation}
Z_Q(w_Q)
=
\frac{\sqrt{\pi}}{2}
e^{Lw_Q/3}
\begin{cases}
\displaystyle
\frac{
\operatorname{erfi}(\sqrt{-L w_Q})
}{
\sqrt{-L w_Q}
},
& w_Q<0,
\\[10pt]
\displaystyle
\frac{
\operatorname{erf}(\sqrt{L w_Q})
}{
\sqrt{L w_Q}
},
& w_Q>0.
\end{cases}
\label{eq:ZQ_rigid}
\end{equation}
The closed form expression for $Z_Q(w_Q)$ enables an explicit parametric construction of $g(S)$ in the rigid-chain limit, with the relation $w_Q^*(S)$ obtained by numerically inverting $S(w_Q)$. The resulting functions $g(S)$, $g'(S)$, and $g(S)-\frac12\alpha S^2$ are shown in Fig.~\ref{fig:g_rigid}(a, b). The function $g(S)$ exhibits singular behavior near the physical bounds $S=1$ and $S=-1/2$. The competition between orientational entropy and alignment interactions produces a nematic minimum at finite $S$ for sufficiently large $\alpha$. As shown in the inset of Fig.~\ref{fig:g_rigid}(b), $g'(S)\sim(1-S)^{-1}$ as $S\to1$, implying a logarithmic asymptotic behavior $g(S)\sim-\ln(1-S).$ This reflects the rapid loss of orientational entropy as the system approaches perfect alignment. Figure~\ref{fig:g_rigid}(c) shows the phase diagram in the rigid-chain limit, including the isotropic-nematic coexistence line (binodals) and the nematic spinodal. The ratio $u_0/u_2$ strongly influences the width of the coexistence region: decreasing $u_0/u_2$ broadens the coexistence window, consistent with previous studies~\cite{qing2025self,jiang2017thermodynamics}. In addition, the location of the nematic spinodal depends on $u_0/u_2$. Figure~\ref{fig:g_rigid}(d) shows the grand potential density for $u_0/u_2=1$ at the coexistence point and at the nematic spinodal. At coexistence, the isotropic and nematic minima have equal grand potential density, corresponding to the binodal shown in Fig.~\ref{fig:g_rigid}(c). Likewise, the nematic spinodal corresponds to the point at which the nematic minimum loses stability, characterized by a vanishing curvature, $\partial^2(\beta\omega)/\partial S^2=0$, at the stationary point $S_N$.

\section{Linear response}
\label{sec:lr}
The effective free energy constructed in the previous section describes bulk thermodynamics of spatially uniform states, but does not characterize the free energy cost of spatial variations about those states. The purpose of this section is to obtain the leading long wavelength corrections. We therefore consider the linear response of a uniform state to weak, spatially varying external fields. The resulting wave vector dependent susceptibilities incorporate leading corrections.

Two classes of long wavelength variations are of particular interest. As reflected in the effective free energy in Eq.~\eqref{eq:eff_f}, spatial variations in the density $\rho$ and the degree of nematic order $S$ are coupled and must therefore be treated through a coupled linear response analysis. Their wave vector dependent response determines the collective bulk modes, and the characteristic length scales over which perturbations in density and nematic order decay toward the uniform state. These correlation lengths will later provide a natural reference measure for interpreting spatial structures and extent of defect cores. Director distortions, in contrast, decouple from variations in $\rho$ and $S$ at linear order. For a small rotation of a uniform uniaxial state, the leading variation of the tensor order parameter $\delta Q_{ij}(\mathbf r)$ is linear in the local director distortion, whereas the accompanying variations in $\rho(\mathbf r)$ and $S(\mathbf r)$ enter only at higher order. The director response can therefore be analyzed independently, with its long-wavelength behavior determining the Frank elastic constants.

\subsection{Nematic elasticity}
\label{subsec:lr_elasticity}
We first focus on the linear response of the director field in a uniaxial nematic phase. In the long wavelength limit, the elastic free energy cost of director distortions is described by the Frank free energy~\cite{selinger2016introduction}
\begin{equation}
\begin{aligned}
F=\int d\mathbf r\,\Bigl[&
\frac12K_1(\nabla\cdot\mathbf n)^2
+\frac12K_2(\mathbf n\cdot\nabla\times\mathbf n)^2\\
&+\frac12K_3 \left| \mathbf n\times(\nabla\times\mathbf n)\right|^2\Bigr].
\end{aligned}
\end{equation}
Here, $K_1$, $K_2$, and $K_3$ are elastic constants corresponding to the splay, twist, and bend elastic modes, respectively. To isolate individual deformation modes, we consider one dimensional director distortions along the $x$-direction and parameterize the director $\mathbf n(x)$ in terms of a single angle field $\theta(x)$, which represents deviations from a reference orientation $\mathbf n_0$. Different deformation modes (splay, twist, and bend) can be realized by appropriate choices of the reference orientation $\mathbf n_0$ and the plane of director variation. Specifically, splay, twist, and bend correspond to director variations in the $xz$, $yz$, and $xz$ planes, with the reference orientation $\mathbf n_0$ along the $z$, $z$, and $x$ directions, respectively. In the small distortion limit, the Frank free energy density for all three deformation modes reduces to a common quadratic form,
\begin{equation}
f_i \simeq \frac{1}{2}K_i\left(\frac{d\theta}{dx}\right)^2,
\qquad i = 1,2,3.
\end{equation}

As a representative example, we consider twist deformation, parameterized by $\mathbf n=(0,\sin\theta,\cos\theta)$ with $|\theta|\ll1$. To excite a pure twist mode in linear response, we introduce an external field $v_\theta(\mathbf r, \mathbf u) = - h_\theta(x)\, u_y u_z $ that couples to the off-diagonal component $Q_{yz}$, with Hamiltonian
\begin{equation}
\beta H_{\text{ext}} = - \int d\mathbf r \, h_\theta(x)\, Q_{yz}(x).
\end{equation}
For small distortions around a uniaxial nematic state, the tensor field $\mathbf Q(x)$ is assumed to remain approximately uniaxial and can be written as $Q_{ij}(x)\simeq \rho(x)S(x)[n_i(x)n_j(x)-\frac13\delta_{ij} ]$. To leading order in the small angle, one obtains $Q_{yz}(x) \simeq \rho(x) S(x)\,\theta(x)$. Expanding about a uniform reference state and retaining only linear order terms in the distortions, we obtain $Q_{yz}(x) \simeq \rho_0 S_0\,\theta(x)$, where $\rho_0$ and $S_0$ denote the density and uniaxial order parameter of the reference state. Variations in $\rho$ and $S$ contribute only at higher order and are therefore neglected at the linear response level. Accordingly, the external field reduces to an effective linear coupling with the rotation angle,
\begin{equation}
\beta H_{\mathrm{ext}}
\simeq
-\int d\mathbf r\, \rho_0 S_0 h_\theta(x)\theta(x).
\end{equation}
For one-dimensional distortions described by $\theta(x)$, the elastic free energy is proportional to $K_2 k^2 \theta(k)^2$ in Fourier space, while the external field provides a linear term proportional to $\rho_0 S_0\, h_\theta(k) \theta(k)$. Here $k = 2\pi/\lambda$ denotes the magnitude of the wavevector, where $\lambda$ is the corresponding distortion wavelength. Minimizing the elastic free energy yields
\begin{equation}
\theta(k) =\chi_{2}(k) h_\theta(k)= \frac{\rho_0 S_0}{\beta K_2 k^2}\, h_\theta(k),
\label{eq:theta_response}
\end{equation}
where $\chi_2$ is the linear susceptibility associated with the twist mode. This expression exhibits the characteristic $k^{-2}$ scaling of a Goldstone mode. The same construction can be applied to the splay and bend modes by choosing the corresponding deformation geometry and probe fields, leading to mode dependent susceptibilities $\chi_i(k) = \rho_0 S_0/(\beta K_i k^2)$ for $i=1,2,3$. Consequently, the inverse susceptibilities are linear in $k^2$ in the long-wavelength limit, with slopes directly determined by the corresponding Frank elastic constants.

\subsection{Density--nematic degree of order coupling}
\label{subsec:lr_coupling}

We now turn to the linear response to density and uniaxial order parameter variations around a spatially uniform reference state. For the density part, consider an orientation independent (isotropic) external field that couples to the local segment density, $v_{\rho}(\mathbf r,\mathbf u) = -h_{\rho}(\mathbf r)$. The corresponding energy is
\begin{equation}
\beta H_{\mathrm{ext}} 
= -\int d\mathbf r \, h_{\rho}(\mathbf r)\, \hat\rho(\mathbf r).
\end{equation}
For the uniaxial order parameter, consider a uniaxial field with easy axis $\mathbf e$, $v(\mathbf r,\mathbf u)=-\,h_Q(\mathbf r)\left[(\mathbf u\cdot\mathbf e)^2-\frac13\right],$ where $h_Q(\mathbf r)$ controls the alignment strength to the external field. The corresponding contribution to the free energy is given by
\begin{equation}
\beta H_{\mathrm{ext}}
=
-\int d\mathbf r \, h_Q(\mathbf r)\,
E_{ij}(\mathbf e)\, Q_{ij}(\mathbf r),
\end{equation}
where $E_{ij}(\mathbf e)=e_i e_j-\frac13\delta_{ij}$ is the traceless uniaxial tensor associated with $\mathbf e$. We define the corresponding scalar projection $Q(\mathbf r)\equiv Q_{ij}(\mathbf r)E_{ij}(\mathbf e)$, so that $h_Q(\mathbf r)$ couples linearly to $Q(\mathbf r)$. Throughout the following analysis, the boldface symbol $\mathbf Q$ denotes the full nematic order tensor, while the non-bold symbol $Q$ denotes the scalar quantity.

For the nematic reference state, we choose the external field to be aligned with the director, $\mathbf e=\mathbf n$, so that the response probes density and uniaxial order parameter variations without mixing in director distortions. While in the isotropic reference state, the director $\mathbf n$ is induced by the external perturbation and coincides with $\mathbf e$. Assuming that the system remains locally uniaxial under small perturbations, we approximate $Q_{ij}(\mathbf r)\simeq \rho(\mathbf r)S(\mathbf r)\left(n_i n_j-\frac13\delta_{ij}\right).$ With this choice, the projection simplifies to $Q(\mathbf r)\simeq \frac{2}{3}\rho(\mathbf r)S(\mathbf r)$, showing that $Q$ represents a density-weighted nematic order parameter.

The response to small field induced deviations from the uniform reference state is analyzed in Fourier space. Introducing the wavevector $\mathbf k$, one defines
\begin{equation}
\delta \mathbf x(\mathbf k)
=
\begin{pmatrix}
\delta \rho(\mathbf k) \\
\delta Q(\mathbf k)
\end{pmatrix},
\qquad
\mathbf h(\mathbf k)
=
\begin{pmatrix}
h_{\rho}(\mathbf k) \\
h_Q(\mathbf k)
\end{pmatrix},
\end{equation}
where $\delta \mathbf x(\mathbf k)$ represents $\rho$ and $Q$ deviations with respect to the reference state, and $\mathbf h(\mathbf k)$ are the corresponding external fields. Within linear response theory, the free energy cost of small long-wavelength deviations from the uniform state can be expressed to quadratic order as
\begin{equation}
\beta F^{(2)}
=
\frac{1}{2}\sum_{\mathbf k}
\delta \mathbf x^{\mathrm T}(-\mathbf k)\,
\mathbf{M}(\mathbf k)\,
\delta \mathbf x(\mathbf k)
-
\sum_{\mathbf k}
\mathbf h^{\mathrm T}(-\mathbf k)\,
\delta \mathbf x(\mathbf k),
\label{eq:pert_f}
\end{equation}
where $\mathbf{M}(\mathbf k)$ is the inverse susceptibility kernel. The off-diagonal terms in $M_{ij}$ quantify the coupling between $\rho$ and $Q$. By construction, $\mathbf M(\mathbf k)$ is symmetric, such that $M_{\rho Q}=M_{Q\rho}$. In the long-wavelength limit (\(k \equiv |\mathbf k|\to 0\)),  $\mathbf{M}(\mathbf k)$ is expanded to quadratic order in $\mathbf k$,
\begin{equation}
M_{ij}(\mathbf k)
=
A_{ij}
+
K_{ij}^{\alpha\beta}k_\alpha k_\beta,
\label{eq:M_small_k}
\end{equation}
where $A_{ij}$ describes the uniform ($k=0$) response and $K_{ij}^{\alpha\beta}$ characterize the leading response to spatial variations, with $i,j\in\{\rho,Q\}$ and $\alpha,\beta\in\{x,y,z\}$. The coefficients $A_{ij}$ play the same role as the Hessian matrix introduced in Sec.~\ref{sec:uniform}, representing the response to uniform variations. Here the variables are chosen as $(\rho,Q)$ rather than $(\rho,S)$, so the individual matrix elements differ numerically. Nevertheless, both formulations describe the same underlying stability of the uniform state and therefore lead to the same spinodal condition, which, for the present two-variable problem, is equivalent to the vanishing of $\det A$, i.e., $A_{\rho\rho} A_{QQ} - A_{\rho Q}^2 = 0$. Importantly, this condition shows that the onset of instability depends on the coupling between $\rho$ and $Q$. Otherwise ($A_{\rho Q}=0$), instability would occur independently in each channel when either $A_{\rho\rho}=0$ or $A_{QQ}=0$. A finite coupling modifies this criterion and can drive the system unstable even when both diagonal terms remain positive. 

According to Eq.~\eqref{eq:eff_f}, the resulting coefficients obtained after transforming variables from $(\rho,S)$ to $(\rho,Q)$, with $S=3Q/(2\rho)$, are
\begin{equation}
\begin{aligned}
A_{\rho\rho}
&=
u_0+\frac{1}{L\rho}
+
\frac{S^2}{L\rho}g'',
\\
A_{\rho Q}
&=
-\frac{3S}{2L\rho}g'',
\\
A_{QQ}
&=
-
\frac{9}{4}u_2+
\frac{9}{4L\rho}g''.
\end{aligned}
\label{eq:A_ij_theory}
\end{equation}
Note the dependence of all three terms on the second derivative of the conformational free energy. Also, only $A_{\rho\rho}$, which describes density fluctuations at fixed $Q$, depends on $u_{0}$.

For a uniaxial reference state with director $\mathbf n_0$, rotational symmetry implies
a decomposition into longitudinal and transverse components,
\begin{equation}
K_{ij}^{\alpha\beta}
=
K_{ij}^{\perp}\delta_{\alpha\beta}
+
\left(
K_{ij}^{\parallel}
-
K_{ij}^{\perp}
\right)
n_{0\alpha}n_{0\beta},
\end{equation}
so that
\begin{equation}
K_{ij}^{\alpha\beta}k_\alpha k_\beta
=
K_{ij}^{\perp}k_\perp^2
+
K_{ij}^{\parallel}k_\parallel^2,
\label{eq:K_perp_para}
\end{equation}
where $ k_\parallel = \mathbf k\cdot\mathbf n_0, \ k_\perp^2= k^2-k_\parallel^2.$ For brevity, labels indicating the perpendicular or parallel direction will be omitted when only one direction is under consideration. 

The linear response follows from minimizing the free energy in Eq.~\eqref{eq:pert_f}, yielding
\begin{equation}
\delta \mathbf x(\mathbf k)=\boldsymbol{\chi}(\mathbf k)\mathbf h(\mathbf k),
\end{equation}
where $\boldsymbol{\chi}(\mathbf k) \equiv  \mathbf{M}^{-1}(\mathbf k)$ is the susceptibility matrix. The matrix elements of the susceptibility $\boldsymbol{\chi}(\mathbf k)$ can be obtained from the linear response to weak external fields $h_\rho(\mathbf k)$ and $h_Q(\mathbf k)$ applied separately (with the other set to zero). The induced variations in $\rho$ and $Q$ define the response amplitudes, whose ratios to the applied field strength yield all four elements of $\boldsymbol{\chi}(\mathbf k)$.

To gain further physical insight, it is useful to analyze the coupled response in terms of the eigenmodes of $\mathbf M(\mathbf k)$. The corresponding eigenvalues are
\begin{equation}
\begin{aligned}
\lambda_{\pm}(\mathbf k)
=\frac{1}{2}\left[
M_{\rho\rho}+M_{QQ} 
\pm \sqrt{
\left(M_{\rho\rho}-M_{QQ}\right)^2+
4M_{\rho Q}^2}
\right].
\end{aligned}
\end{equation}
The collective behavior is dominated by the mode with the smaller eigenvalue, $\lambda_{-}(\mathbf k)$, referred to as the soft mode. Expanding the soft mode eigenvalue in the long-wavelength limit gives
\begin{equation}
\lambda_{-}(\mathbf k)
\approx
A_{-}
+
K_{-}^{\perp}k_\perp^2
+
K_{-}^{\parallel}k_\parallel^2 .
\end{equation}
The associated longitudinal and transverse correlation lengths are defined as $\xi_-^{\parallel,\perp}=\sqrt{K_-^{\parallel,\perp}/A_-}.$ The correlation length $\xi_-^{\parallel,\perp}$ characterizes the spatial scale of the soft mode, governing the decay of the coupled modes. 

While the soft mode characterizes the collective responses of the coupled system, the quantity most directly accessed in simulations and experiments is $Q$. It is therefore also useful to define a corresponding correlation length associated with this observable. We consider the projected response in the $Q$ channel by minimizing with respect to $\rho$, which yields an effective susceptibility $\chi_{Q}(\mathbf k) = \left[\mathbf{M}^{-1}(\mathbf k)\right]_{QQ}$. Correspondingly, its inverse takes the form
\begin{equation}
\chi_Q^{-1}(\mathbf k) = M_{QQ}(\mathbf k) -\frac{ M_{\rho Q}^2(\mathbf k)}{M_{\rho\rho}(\mathbf k)}.
\end{equation}
In the long-wavelength limit, it can be expanded as $\chi_{Q}^{-1}(\mathbf k)\approx A_Q+K_{Q}^{\perp}k_\perp^2
+K_{Q}^{\parallel}k_\parallel^2,$ thereby defining effective coefficients $A_{Q}$ and $K_{Q}^{\parallel, \perp}$. The corresponding correlation length is then given by $\xi_Q^{\parallel, \perp} \equiv \sqrt{K_Q^{\parallel, \perp} / A_Q}$. 

\section{Results and Discussion}
\label{sec:results}
\begin{figure*}[!t]
\centering
    \includegraphics[width=0.95\textwidth]{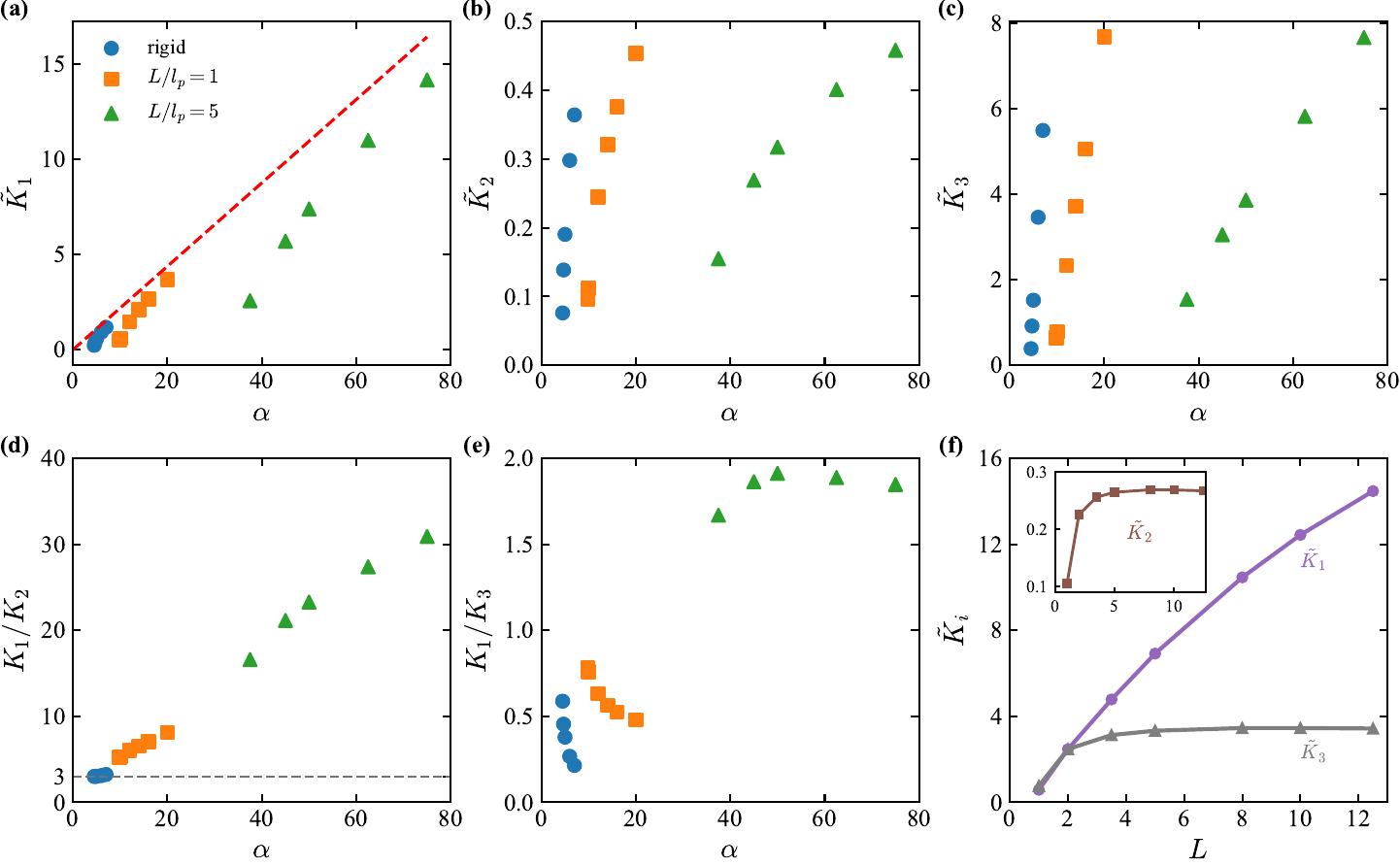}
\caption{
Scaled Frank elastic constants $\widetilde K_i$ and their ratios as functions of $\alpha$ and chain length $L$. Panels (a)--(c) show $\widetilde K_1$, $\widetilde K_2$, and $\widetilde K_3$, respectively, as functions of $\alpha$ for different chain flexibilities. Panel (d) shows the ratio $K_1/K_2$, and panel (e) the ratio $K_1/K_3$. Blue circles, orange squares, and green triangles correspond to rigid chains, semiflexible chains with $L/l_p=1$, and $L/l_p=5$, respectively. In panel (a), the dashed line indicates the asymptotic scaling in the Onsager limit for strongly aligned rigid rods. In panel (d), the dashed line marks the theoretical rigid-chain relation $K_1/K_2=3$. Panel (f) shows the chain-length dependence of $\widetilde K_i$ at fixed segment density $nL/V=1$ and persistence length $l_p=1$.
}
\label{fig:Ki_alpha}
\end{figure*}

We introduce dimensionless variables as follows: Energetic quantities are rendered dimensionless by multiplication with $\beta=1/(k_BT)$, as in $\beta F$ and $\beta H$. Length-related quantities are measured relative to a reference length $L_0$, corresponding to $L\to L/L_0$, $l_p\to l_p/L_0$, $\mathbf r\to\mathbf r/L_0$, and $u_{0,2}\to u_{0,2}/L_0$. The Frank elastic constants are reported as $\beta K_iL_0$. In the following, only dimensionless variables are reported. Unless otherwise stated, all numerical results are obtained from SCFT calculations by using a real spherical harmonics representation. The spherical harmonic expansion is truncated at degree $l_{\max}=4$, except for the Frank elastic constants in Fig.~\ref{fig:Ki_alpha}, for which $l_{\max}=6$ is used to improve numerical accuracy. Details of the numerical implementation are provided in Appendix~\ref{appendix:numerics}.

We first mention results for the isotropic-nematic transition at fixed density~\cite{qing2025self,spencer2020nematic} to establish the notation and physical context for the following analysis. For a uniform state at fixed density, the equilibrium nematic order parameter is determined by the stationarity condition Eq.~\eqref{eq:extrema}. A nematic solution exists only when $\alpha>\alpha^\ast(L/l_p)$, where the threshold effective alignment strength $\alpha^\ast$ depends on chain flexibility through $L/l_p$. The corresponding nematic state is characterized by $S_N=S_N(\alpha,L/l_p).$ Within the nematic regime, numerical results show that $S_N$ increases monotonically with $\alpha$ at fixed $L/l_p$, whereas at fixed $\alpha$ it decreases with increasing $L/l_p$.

\subsection{Director response and Frank elastic constants}

Having formulated the director linear response analysis in Sec.~\ref{subsec:lr_elasticity}, we compute the Frank elastic constants from SCFT. Numerical results show that $\chi_i^{-1}(k)$ is linear in $k^2$ with a negligible intercept over the range considered, consistent with the Goldstone-mode behavior in Eq.~\eqref{eq:theta_response}. The Frank elastic constant for each mode is then obtained directly from the corresponding slope. We further find that the induced variations in $\rho$ and $S$ contain no component at wavevector $k$, appearing only at $2k$. Thus, within linear response, the director modes are effectively decoupled from density and nematic order variations. 

\subsubsection{Scaling and elastic anisotropy}
The elastic constants obtained are shown in Fig. \ref{fig:Ki_alpha}. A scaling form
\begin{equation}
\beta K_i
=
\frac{1}{u_2}
\tilde K_i(\alpha,L/l_p),
\label{eq:K_tilde}
\end{equation}
can be introduced, where $\tilde K_i$ is a scaling function. Data obtained for different values of $L$, $u_2$, and $n/V$ collapse onto this representation. In contrast, varying $u_0$ has negligible effect on $\beta K_i$.

This naturally raises the question of how $\tilde K_i$ depends on $\alpha$ and $L/l_p$. We examine this dependence for representative values $L/l_p=0,\,1,\,5$, spanning the rigid-chain limit to increasingly flexible chains. The results are summarized in Fig.~\ref{fig:Ki_alpha}: panels (a)--(c) show $\tilde K_1$, $\tilde K_2$, and $\tilde K_3$, respectively, while panels (d) and (e) show the ratios $K_1/ K_2$ and $ K_1/ K_3$. The data cover the nematic order range $S_N\sim0.4$--$0.7$ for each $L/l_p$. The relative magnitudes of the elastic constants change substantially with chain flexibility. In the rigid-chain limit, $\tilde K_1$ and $\tilde K_2$ are of the same order, whereas increasing flexibility strongly suppresses $\tilde K_2$, so that $\tilde K_1$ exceeds $\tilde K_2$ by approximately one order of magnitude. A similar but more striking crossover occurs between splay and bend: $\tilde K_1<\tilde K_3$ for rigid and moderately flexible chains, while $\tilde K_1>\tilde K_3$ for $L/l_p=5$. These trends show that chain flexibility affects twist and bend much more strongly than splay. To further examine this contrasting flexibility dependence, we vary the chain length $L$ at fixed $l_p$ while keeping the contour length density $nL/V$ fixed, allowing direct comparison with experiments at fixed volume fraction. The results are shown in Fig.~\ref{fig:Ki_alpha}(f). Over the range considered, $\tilde K_1$ increases monotonically with $L$, whereas $\tilde K_2$ and $\tilde K_3$ initially increase and then approach plateaus at larger $L$. Thus, as $L/l_p$ increases at fixed $l_p$, the twist and bend elastic constants become progressively less sensitive to the total chain length and are instead controlled primarily by the persistence length $l_p$.

We note that Frank elastic constants for wormlike chain models have also been studied using an alternative approach based on a fluctuation free energy~\cite{ghosh2022semiflexible}. In that work, the MS interaction strength was used as the primary control parameter. Through the scaling form in Eq.~\ref{eq:K_tilde}, those results can be related to the present formulation.

\subsubsection{Rigid chain limit and asymptotic scaling}
To further understand these trends, we consider the rigid chain limit and compare our results with the analytical prediction in the Onsager model for strongly aligned rigid rods~\cite{odijk1986elastic},
\begin{equation}
\beta K_1 = 3 \beta K_2 =\frac{7}{32} \frac{1}{D} \left( \frac{n}{V} L^2 D \right), 
\;
\beta K_3 =\frac{\pi}{48}  \frac{1}{D} \left( \frac{n}{V} L^2 D \right)^3,
\label{eq:odijk_rigid}
\end{equation}
where $D$ is the rod diameter. Although these expressions are derived for excluded volume interactions between rigid rods, retaining only the $P_2(\mathbf u\cdot\mathbf u')$ contribution to the orientational dependence of the Onsager excluded volume kernel gives the same quadrupolar form as the MS interaction. The coefficient of this term is proportional to the rod diameter $D$, so that $D$ and $u_2$ play similar roles. Within the correspondence $u_2 \sim D$, one has $\alpha \sim \Phi (L/D)$, where $\Phi$ is the volume fraction. For fixed $L/l_p$ and $L/D$, varying $\alpha$ is therefore equivalent, at the scaling level, to scanning the volume fraction. Accordingly, in the high-$S$ asymptotic regime, Eq.~\eqref{eq:odijk_rigid} predicts $\tilde K_1, \tilde K_2 \sim \alpha$, while $\tilde K_3 \sim \alpha^3$. 

Unlike the asymptotic power laws above, the exact rigid rod relation $\widetilde K_1=3\widetilde K_2$ does not require the strong-alignment limit~\cite{odijk1986elastic}. Our numerical results approximately reproduce this relation, with deviations arising from the finite truncation of the real spherical harmonic expansion. Increasing the truncation order from $l_{\max}=4$ to $l_{\max}=6$ substantially reduces these deviations, bringing $K_1/K_2$ close to $3$ over the range considered.

For comparison, Fig.~\ref{fig:Ki_alpha}(a) includes the asymptotic prediction for $\tilde K_1$ from Eq.~\eqref{eq:odijk_rigid}. Although our results are obtained for the MS model in the intermediate ordering regime, they already approach this asymptotic form reasonably well. A more stringent test of the large $S_N$ behavior would require access to stronger ordering, either through a substantially higher spherical harmonic truncation or through an alternative, more accurate treatment of the rigid chain limit. 

For $\tilde K_3$, the rigid-chain results increase more rapidly with $\alpha$ than $\tilde K_1$ and $\tilde K_2$, as reflected in the decreasing $K_1/K_3$ shown in Fig.~\ref{fig:Ki_alpha}(e). This is qualitatively consistent with the stronger growth of the bend elasticity predicted in Eq.~\eqref{eq:odijk_rigid}. However, the present intermediate-ordering data are insufficient to determine whether the same asymptotic $\alpha^3$ scaling carries over to the current MS model.

\subsubsection{Semiflexible chain scaling}
We next examine how finite chain flexibility modifies elastic anisotropy. Our numerical results show that increasing $L/l_p$ suppresses $\tilde K_2$ and $\tilde K_3$ much more strongly than $\tilde K_1$. This contrasting sensitivity is consistent with established physical pictures for semiflexible nematics. The splay elasticity is closely related to the density of chain ends~\cite{ciferri1982polymer}; at fixed chain length $L$, varying $l_p$ leaves the end density unchanged, providing a natural explanation for the comparatively weak dependence of $\tilde K_1$ on chain flexibility. Over the range considered, $\tilde K_1$ increases with $\alpha$ in a qualitatively similar manner for all $L/l_p$, while remaining only weakly affected by chain flexibility. By contrast, in the strongly ordered and highly flexible limit, $S_N\to1$ and $L/l_p\gg1$, deflection-length arguments predict~\cite{odijk1986elastic}
\begin{equation}
\beta K_2 \sim \frac{1}{D} \left( \frac{n}{V} L^2 D \frac{l_p}{L} \right)^{1/3} , 
\quad
\beta K_3 \sim \frac{1}{D} \left( \frac{n}{V} L^2 D \frac{l_p}{L}\right).
\label{eq:semi_scaling}
\end{equation}
Using the correspondence $u_2 \sim D$, these relations can be expressed in terms of $\alpha$ as $\tilde K_2 \sim ( \alpha l_p / L )^{1/3}$ and $\tilde K_3 \sim \alpha l_p / L$. As chains become more flexible, the ratio $l_p/L$ decreases, suppressing the growth of $\tilde K_2$ and $\tilde K_3$ with increasing $\alpha$, consistent with our numerical results. Furthermore, expressing the scaling in terms of the volume fraction $\Phi$, one obtains $\tilde K_2 \sim (\Phi l_p/D)^{1/3}$ and $\tilde K_3 \sim \Phi l_p/D$. These relations suggest that, for sufficiently flexible chains at fixed $\Phi$, both $\tilde K_2$ and $\tilde K_3$ are governed primarily by $l_p$ rather than $L$, consistent with the crossover observed in Fig.~\ref{fig:Ki_alpha}(f) near $L/l_p\gtrsim5$.

\begin{figure*}[!t]
\centering
\includegraphics[width=0.72\linewidth]{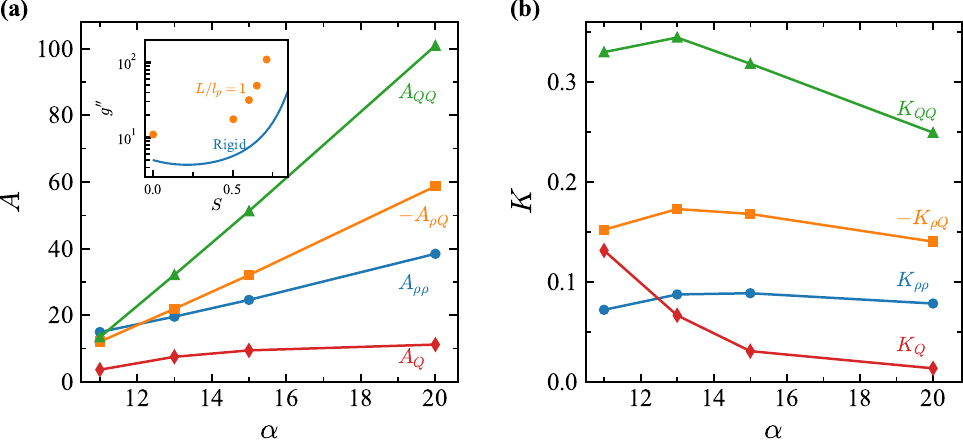}
\caption{
Dependence of \(A_{ij}\) and \(K_{ij}\) on \(\alpha\) for $L=l_p=1$, $u_0=10$, $u_2=10$ and $\mathbf k\perp\mathbf n$, with $\alpha$ varied through $n/V$. (a) \(A_{\rho\rho}\), \(-A_{\rho Q}\), \(A_{QQ}\), and \(A_Q\) versus \(\alpha\). Inset: comparison of \(g''(S)\) for the rigid-chain limit and the finite-stiffness case \(L/l_p=1\) as a function of \(S\), where the \(L/l_p=1\) data are extracted from \(A_{ij}\). The point at $S=0$ is obtained from the isotropic reference state and gives the corresponding value of $g''(0)$. (b) \(K_{\rho\rho}\), \(-K_{\rho Q}\), \(K_{QQ}\), and \(K_Q\) versus \(\alpha\).
}
\label{fig:A_K_vs_alpha}
\end{figure*}

\subsubsection{Scope of the asymptotic comparison}
Equations~\eqref{eq:odijk_rigid} and \eqref{eq:semi_scaling} are Onsager-based asymptotic results valid in the strongly ordered limit, $S_N\to1$. We use these results primarily as qualitative benchmarks for the relative behavior of the three elastic modes, rather than as predictions of the power law exponents in the present model. In particular, they capture two trends also observed in our numerical results: the stronger growth of the bend elasticity in the rigid chain limit, and the enhanced sensitivity of twist and bend to chain flexibility in the semiflexible regime. Our calculations at intermediate order establish these trends as $\alpha$ is increased, while the corresponding asymptotic power laws are expected to emerge only at much larger $S_N$. The exponents may also depend on the form of the orientational interaction, so the MS model considered here need not exhibit exactly the same scaling as the Onsager theory. Testing the asymptotic powers would require calculations much closer to $S_N=1$ and correspondingly higher spherical-harmonic truncations, which are beyond the scope of the present work.

\begin{figure*}[t]
\centering
\includegraphics[width=0.72\textwidth]{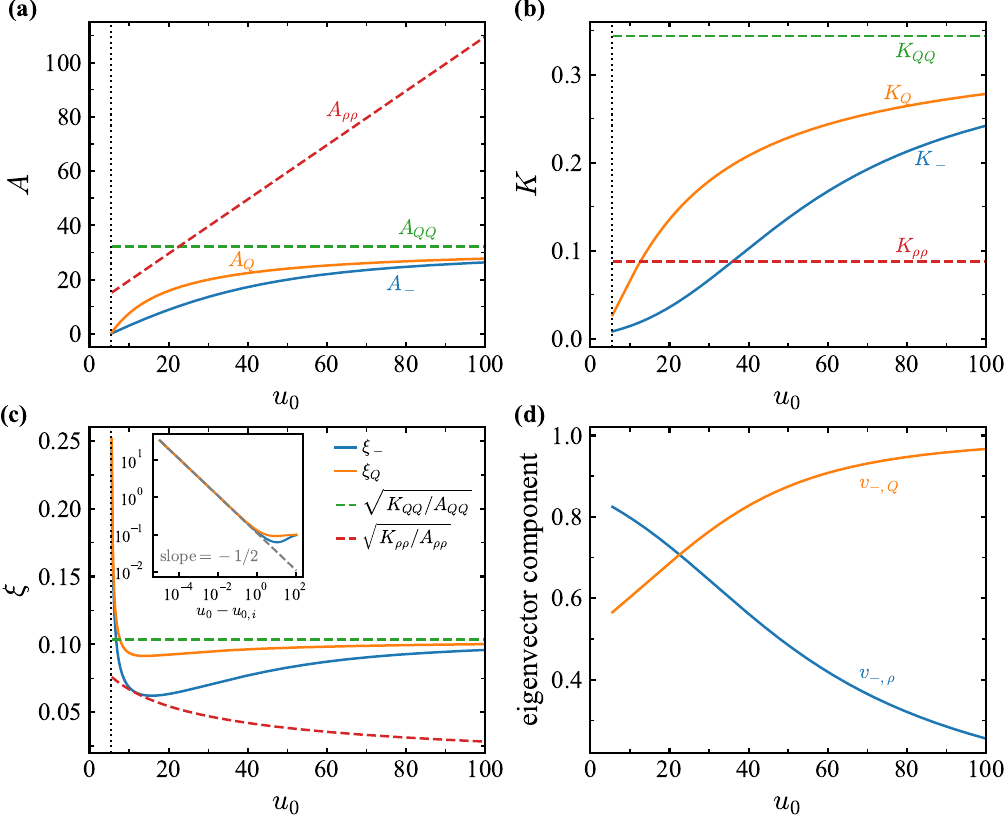}
\caption{
Dependence on $u_0$ of (a) the $k=0$ coefficients, (b) the gradient coefficients, (c) the correlation lengths, and (d) the soft mode eigenvector components.  The coefficients are constructed using the $A_{ij}$ and $K_{ij}$ values extracted from a $\mathbf k\perp\mathbf n$ nematic reference state with $L=l_p=1$, $n/V=1.3$, and $u_2=10$. All coefficients are then held fixed at their reference state values except for $A_{\rho\rho}$, which varies with $u_0$ as $A_{\rho\rho}=u_0+9.586$. The vertical dashed line in (a, b, c) denotes the instability point $u_0 = u_{0,i}$, at which $A_Q= A_- = 0$. In panel (c), the inset shows a log-log plot of $\xi$ versus $u_0 - u_{0,i}$, showing the scaling $\xi \sim (u_0 - u_{0,i})^{-1/2}$ near the instability point.
}
\label{fig:u0_scan}
\end{figure*}
\subsection{Density--nematic order coupling and correlation lengths}
Having formulated the coupled density--nematic order linear response in Sec.~\ref{subsec:lr_coupling}, we now examine the numerical results for the long-wavelength response and the associated correlation lengths. For both isotropic and
nematic reference states, the numerically extracted inverse susceptibility $\mathbf M(\mathbf k)=\boldsymbol{\chi}^{-1}(\mathbf k)$ is symmetric within numerical accuracy and exhibits the expected quadratic dependence on wavevector. Representative fits of the matrix elements of $\mathbf{M}$ in the isotropic state and in the nematic $\mathbf k\parallel \mathbf n$ and $\mathbf k\perp \mathbf n$ geometries are shown in Appendix~\ref{app:lr_details}, Fig.~\ref{fig:M_elements_q2_app}.

The key observation relevant to the nematic phase is that density and nematic order are strongly coupled. The off-diagonal coefficients $A_{\rho Q}$ and $K_{\rho Q}$ are comparable in magnitude to the corresponding diagonal terms and therefore cannot, in general, be neglected when determining the collective response. In addition, the gradient coefficients are anisotropic, with distinct $K_{ij}^{\parallel}$ and $K_{ij}^{\perp}$ for variations parallel and perpendicular to the equilibrium director, whereas the local coefficients $A_{ij}$ are independent of direction. As a consequence of the full coupled response, the correlation lengths are also anisotropic, with $\xi_Q^{\parallel}>\xi_Q^{\perp}$ over the parameter range considered.

To characterize the consequences of density--nematic order coupling in the nematic phase, we first vary $\alpha$ to determine how the local and gradient response coefficients, and hence the correlation lengths, evolve with increasing nematic order. We then vary $u_0$, which directly controls the density response while leaving the equilibrium nematic order unchanged, providing a complementary way to examine how density fluctuations enter the coupled $\rho$--$Q$ response.

\subsubsection{Dependence on nematic ordering strength}
We first examine how the coupled response evolves with the nematic ordering strength $\alpha$ in the nematic reference state. Using $1/(L\rho)=u_2/\alpha$, Eq.~\eqref{eq:A_ij_theory} can be rewritten as
\begin{equation}
\begin{aligned}
A_{\rho\rho}
&=u_0+u_2
\left[
\frac{1}{\alpha}+
\frac{S^2}{\alpha}g''
\right],
\\
A_{\rho Q}
&=-u_2
\left[
\frac{3S}{2\alpha}g''
\right],
\\
A_{QQ}
&=u_2
\left[
-\frac94+
\frac{9}{4\alpha}g''
\right].
\end{aligned}
\label{eq:A_nematic}
\end{equation}
Since $S=S_N(\alpha,L/l_p)$ and $g''=g''(S_N;L/l_p)$, all terms inside the brackets are functions only of $\alpha$ and $L/l_p$. This scaling provides a simple interpretation of the large $\alpha$, rigid-chain limit. In this regime, $g'(S)\sim(1-S)^{-1}$ near $S=1$. Combining this asymptotic form with the stationary condition $g'(S_N)=\alpha S_N$ gives $(1-S_N)^{-1}\sim\alpha$, and therefore $g''(S_N)\sim(1-S_N)^{-2}\sim\alpha^2$. As a result, all $A_{ij}$ exhibit the same leading asymptotic scaling, $A_{ij} \sim u_2\alpha$.

As a representative finite stiffness case, numerical results are presented for the nematic phase at $L/l_p=1$ and $\mathbf k\perp\mathbf n$, with $\alpha$ varied by changing $n/V$. Figure~\ref{fig:A_K_vs_alpha}(a) shows the resulting $A_{ij}$ as functions of $\alpha$. In the explored range corresponding to \(S_N\simeq0.5\)--0.7, the coefficients \(A_{ij}\) display an approximately linear dependence on \(\alpha\). Using Eq.~\eqref{eq:A_nematic}, the quantity $g''(S)$ can be extracted independently from each element of $A_{ij}$. The resulting values from different elements are found to be consistent within numerical accuracy and are shown in the inset of Fig.~\ref{fig:A_K_vs_alpha}(a), together with the rigid-chain result for comparison. The finite stiffness case (\(L/l_p=1\)) exhibits systematically larger values of \(g''\) at the same \(S\), indicating a stronger entropic penalty for nematic ordering. Physically, semiflexible chains possess many local orientational degrees of freedom absent in rigid chains. Increasing nematic order suppresses orientational fluctuations along the chain contour, leading to a substantially larger entropy loss than in the rigid-chain limit. 

For the gradient coefficients $K_{ij}$, we perform numerical scans along different paths in parameter space while keeping $\alpha$ and $L/l_p$ fixed. We find that the numerical data are consistent with the scaling form
\begin{equation}
K_{ij}=u_2L^2\tilde{K}_{ij}(\alpha,L/l_p),
\end{equation}
where $\tilde{K}_{ij}$ is a scaling function depending only on $\alpha$ and $L/l_p$. In contrast to $A_{ij}$, $K_{ij}$ show only weak dependence on $\alpha$, as shown in Fig.~\ref{fig:A_K_vs_alpha}(b). 

To illustrate the effect of the $\rho-Q$ coupling in the nematic phase, the corresponding effective coefficients are also shown in Fig.~\ref{fig:A_K_vs_alpha}. As $\alpha$ increases, $A_Q$ initially rises and then saturates, whereas $K_Q$ decreases significantly. Consequently, the correlation length $\xi_Q$ decreases as the system moves deeper into the nematic phase.

\begin{figure*}[t]
\centering
\includegraphics[width=0.95\textwidth]{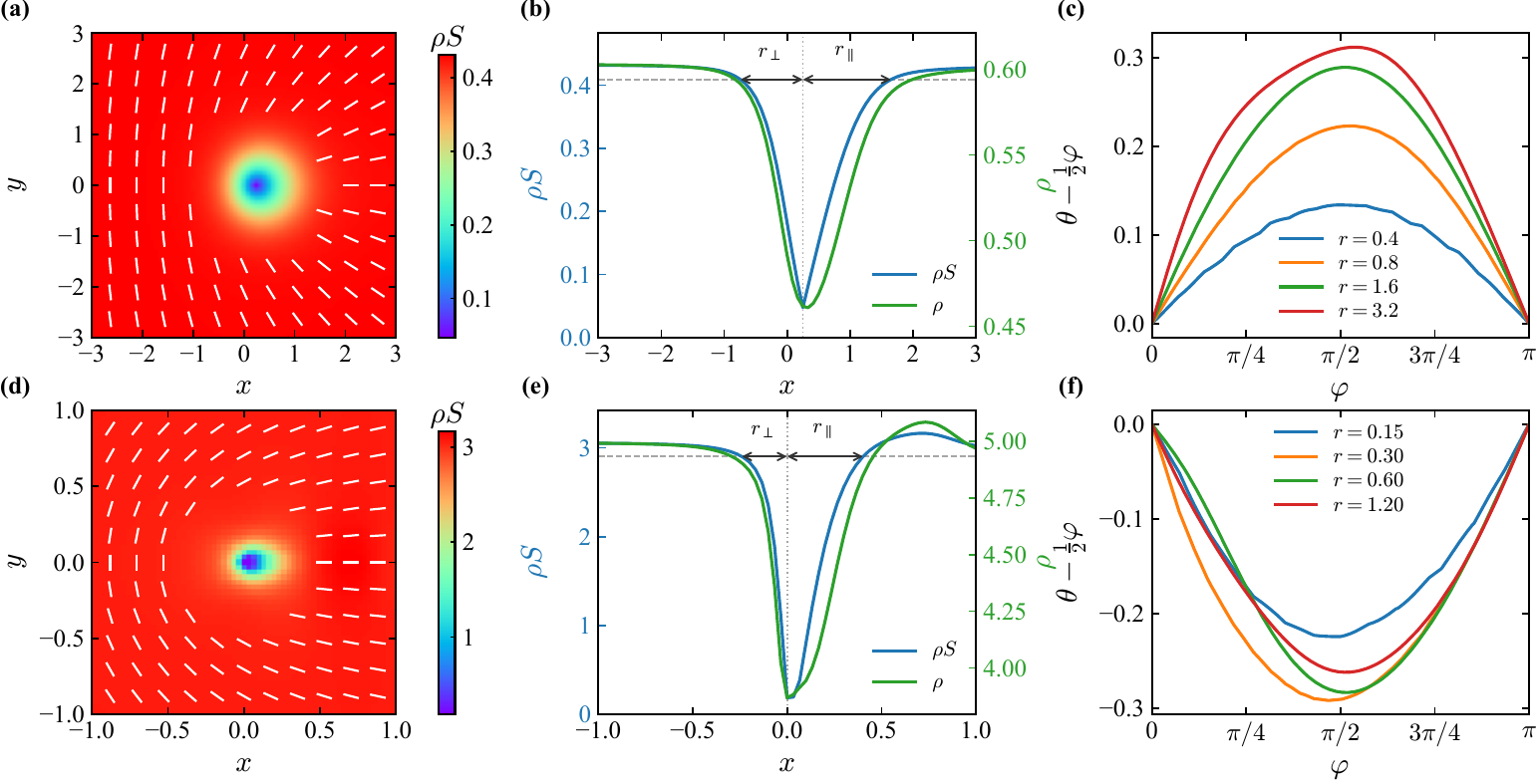}
\caption{Comparison of defect structures in the rigid and semiflexible cases. Top row: rigid case with $L=1$, $l_p \to \infty$ and $n/V=0.6$. Bottom row: semiflexible case with $L=1$, $l_p=0.2$, and $n/V=5$. Both cases use $u_0=u_2=10$. Panels (a) and (d) show the spatial distribution of $\rho S$, with
the white line segments indicating the local director orientation. Panels (b) and (e) show the profiles of $\rho S$ (left axis) and $\rho$ (right axis) along $y=0$. The horizontal dashed lines indicate the $95\%$ recovery level used to determine $r_\perp$ and $r_\parallel$. Panels (c) and (f) show the angular deviation $\theta-\varphi/2$ from the one-constant director profile at several distances $r$ from the defect center.
}
\label{fig:defect_compare}
\end{figure*}

\subsubsection{Approach to the nematic spinodal}
As shown in Sec.~\ref{sec:uniform}, the location of the nematic spinodal depends on the ratio $u_0/u_2$. At fixed $u_2$, varying $u_0$ therefore provides a direct way to tune the nematic reference state toward this instability and to examine how the coupled density--nematic response evolves in its vicinity. In the isotropic reference state, the coupling term vanishes ($A_{\rho Q}=0$), so varying $u_0$ only modifies the coefficient $A_{\rho\rho}$, and leaves the uniform $Q$ response unchanged. In contrast, in the nematic reference state, $A_{\rho Q} \neq 0$. As a result, although varying $u_0$ directly modifies only $A_{\rho\rho}$ and leaves $S_N$ unchanged, it can still destabilize the uniform nematic state by softening a coupled $\rho-Q$ mode. 

As shown in Fig.~\ref{fig:u0_scan}, for the nematic reference state the $\rho$--$Q$ coupling reduces both $A_-$ and $A_Q$ relative to $A_{\rho\rho}$ and $A_{QQ}$. As $u_0$ increases, both $A_-$ and $A_Q$ approach $A_{QQ}$, while the corresponding gradient coefficients approach $K_{QQ}$, reflecting the suppression of density fluctuations and the evolution of the soft mode toward a pure $Q$ mode. Conversely, both $A_-$ and $A_Q$ vanish at a threshold value $u_{0,i}$, corresponding to the nematic spinodal identified in Sec.~\ref{sec:uniform}, whereas the corresponding gradient coefficients remain finite, leading to the divergence of the correlation lengths shown in Fig.~\ref{fig:u0_scan}(c). The inset of Fig.~\ref{fig:u0_scan}(c) shows that the two correlation lengths $\xi_-$ and $\xi_Q$ follow the same scaling $\xi \sim (u_0 - u_{0,i})^{-1/2}$, reflecting the fact that $A_-$ and $A_Q$ vanish linearly with $u_0$ near the spinodal. We next examine the $k=0$ soft mode eigenvector, $\mathbf{v}_-=(v_{-,\rho},v_{-,Q})$, shown in Fig.~\ref{fig:u0_scan}(d). The positive ratio $v_{-,\rho}/v_{-,Q}$ indicates positively correlated $\rho$ and $Q$ variations. As $u_0$ increases, the $\rho$ component is progressively suppressed, causing the soft mode to become increasingly dominated by nematic order variations.

\subsection{Topological defects: Core size and elastic anisotropy}
The results outlined so far indicate that a Landau-de Gennes description needs to be supplemented by a coupling to concentration fluctuations in order to describe lyotropic liquid crystals. A simple coupling scheme is outlined in Appendix \ref{app:ldg_core}. The coupling involves the degree of nematic order, $S$, but not (to lowest order) the director orientation. It is natural to ask how this coupling manifests itself in regions of large order and density gradients (such as regions around disclination cores), beyond the weak distortions analyzed thus far in linear response. Two topics are discussed next: the relation between bulk correlation lengths and core size, and the connection between Frank elastic anisotropy and the director profile.

Linear response predicts anisotropic correlation lengths in the nematic phase, with $\xi_Q^\perp<\xi_Q^\parallel$, consistent with the larger width of isotropic--nematic interfaces for gradients parallel to the director than for perpendicular gradients~\cite{zhou2017fine,popa1997statics,jiang2010isotropic,qing2025self}. Similar anisotropy is observed in $+1/2$ disclination cores~\cite{kim2013morphogenesis,zhou2017fine}. We therefore compare the recovery of nematic order near the core with the corresponding bulk correlation lengths. We also examine whether the anisotropic director profile around a $+1/2$ defect is consistent with the splay--bend anisotropy inferred from the Frank elastic constants.

Two representative $+1/2$ defects which exhibit markedly different elastic anisotropies are considered: one corresponding to rigid chains ($l_p \to \infty$), the other to semiflexible chains with $L/l_p=5$. The SCFT equations are solved numerically as described in Appendix~\ref{appendix:numerics}, and the corresponding $\rho S$ configurations are shown in Fig.~\ref{fig:defect_compare}(a, d). The core radius is estimated from the $\rho S$ profile as the distance from the defect center, identified by the minimum of $\rho S$, to the point at which $\rho S$ recovers to $95\%$ of its bulk value. The corresponding profiles of $\rho$ and $\rho S$ along $y=0$ are shown in Fig.~\ref{fig:defect_compare}(b,e). The core radii measured along directions for which the dominant gradient is locally parallel and perpendicular to $\mathbf n$ are denoted by $r_\parallel$ and $r_\perp$, respectively. With this definition, the core anisotropy is relatively weak in the rigid-chain limit, whereas for $L/l_p=5$ it becomes much more pronounced, with $r_\parallel$ significantly larger than $r_\perp$.

Taking the defect center as the origin, we introduce polar coordinates $(r,\varphi)$, where $\varphi$ is the polar angle and $r$ is the radial distance. We denote by $\theta(r,\varphi)$ the director angle at position $(r,\varphi)$, measured relative to the $\varphi=0$ axis. We use the same notation $\theta$ as in the previous section, the meaning should be clear from the context. For equal splay and bend elastic constants, $K_1=K_3$ (or one-constant approximation), the director field is given by $\theta(r,\varphi)=\varphi/2$. For a $+1/2$ defect in an elastically anisotropic medium, a perturbative expansion to the Dzyaloshinskii solution \cite{dzyaloshinski1970theory,  zhou2017fine, myers2024computational} in small anisotropy gives,
\begin{equation}
\theta(r,\varphi)=\frac{\varphi}{2}+ \frac{3}{4} \epsilon \sin\varphi+ \mathcal{O}(\epsilon^2),
\label{eq:dzya}
\end{equation}
where the elastic anisotropy parameter $\epsilon=(K_3-K_1)/(K_1+K_3)$. The profile we observe deviates from this prediction. Qualitative agreement is found in the approximate sinusoidal dependence shown in Figs.~\ref{fig:defect_compare}(c) and (f). The sign of the deviation is also consistent with Eq. \eqref{eq:dzya}, positive in Fig.~\ref{fig:defect_compare}(c) for a rigid chain ($K_1 < K_3$), and negative in Fig.~\ref{fig:defect_compare}(f) for a semiflexible chain ($K_1 > K_3$).  For the rigid chain case, the SCFT linear response analysis yields $\epsilon \approx 0.5$ under the same parameters, while for $L/l_p = 5$ we find $\epsilon \approx -0.36$. However, the amplitude of the deviation from $\varphi/2$ depends  on the radial coordinate, and it generally decreases as one approaches the defect center. This dependence is not consistent with the Dzyaloshinskii solution, and although qualitatively consistent with a reduction of elastic anisotropy near the core, where both $\rho$ and $S$ are reduced, the amplitude does not scale with the known reduction in elastic constants predicted by the standard map between Frank elastic constants and elastic gradient coefficients in the LdG theory.

\begin{figure}[htbp]
\centering
\includegraphics[width=0.8\columnwidth]{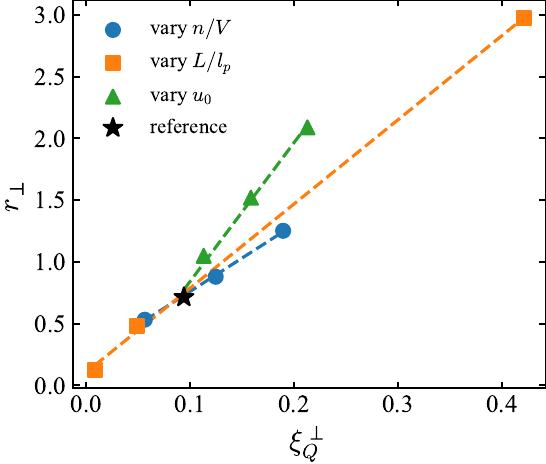}
\caption{
Core radius $r_\perp$ as a function of the correlation length $\xi_Q^\perp$ for parameter scans over $n/V$, $L/l_p$, and $u_0$. Circles, squares, and triangles denote the three parameter scans, respectively, while the dashed lines are linear fits, highlighting the approximately linear relation between $r_\perp$ and $\xi_Q^\perp$. Increasing either $u_0$ or $n/V$ reduces $\xi_Q^\perp$. In the $L/l_p$ scan (with $L=1$ fixed), increasing $l_p$ leads to a larger $\xi_Q^\perp$. The reference parameters are $u_0=u_2=10$, $n/V=1.3$, and $L=l_p=1$. The $u_0$ scan uses $u_0=5.7$, $6$, and $7$. The chain number density scan uses $n/V=1.1$, $1.2$, and $1.5$. For the chain stiffness scan, $L=1$ is fixed and $l_p=0.1$ ($n/V=9.2$), $0.5$ ($n/V=2.2$), and $\infty$ ($n/V=0.5$) are chosen, with $n/V$ adjusted to maintain $S_N\approx0.6$.
}
\label{fig:r_xi}
\end{figure}

We finally compare the disclination core size with the bulk correlation length obtained from linear response, using the perpendicular direction as a representative case. As shown in Fig.~\ref{fig:r_xi}, $r_\perp$ varies approximately linearly with $\xi_Q^\perp$ over a range of $n/V$, $L/l_p$, and $u_0$. The bulk correlation length is therefore a useful measure of the characteristic defect size despite being extracted from small fluctuations about the uniform nematic state. A similar approximately linear relation between the core size and the bulk correlation length is found in the density coupled LdG model presented in Appendix~\ref{app:ldg_core}. Over the parameter range considered here, the defect-core size remains comparable to the chain length $L$, unlike the much larger disparity between core size and mean aggregate length observed in lyotropic chromonics~\cite{re:zhou17b,zhou2017fine}. Thus, density coupling in a system of semiflexible chains is not sufficient to account for the large disparity between core radius and aggregate length observed in chromonics.

The phenomenological extension of the LdG model of Appendix~\ref{app:ldg_core} further allows the defect solution to be followed across a broader parameter range, including the crossover from a localized core toward an isotropic domain inclusion, or tactoid-like, configuration. As this crossover is approached, the approximately linear relation between core size and the bulk correlation length begins to break down. However, as long as the defect remains sufficiently localized, the core size continues to track the correlation length reasonably well.

\section{Conclusion}
\label{sec:conclusion}

We have investigated elastic response, density--nematic order coupling, correlation lengths, and disclination structure in a semiflexible nematic within a wormlike chain SCFT framework. The alignment interaction in the Hamiltonian is of the Maier--Saupe form. In order to model a lyotropic liquid crystal, solvent effects have been incorporated implicitly through an effective repulsive interaction independent of orientation deriving from a Flory--Huggins type interaction between chain segments and solvent molecules.

For uniform equilibrium states, the theory yields an effective free energy in terms of the segment density $\rho$ and uniaxial nematic order parameter $S$, explicitly capturing their coupling. This free energy describes isotropic--nematic coexistence and spinodal instabilities while preserving the physical range $-1/2\le S\le1$ and the associated singular behavior at the boundaries of the domain. By contrast, a conventional low order Landau--de Gennes expansion about $S=0$ does not impose these physical bounds on the order parameter.

Linear response calculations within SCFT yield the Frank elastic constants and the coupled density--nematic susceptibilities. Director distortions remain effectively decoupled from density and nematic order fluctuations at linear order, whereas density and nematic order are strongly coupled in the nematic phase. The elastic constants obey the scaling form $\beta K_i=u_2^{-1}\tilde K_i(\alpha,L/l_p)$, with $\alpha=nL^2u_2/V$. Increasing chain flexibility strongly suppresses twist and bend while affecting splay more weakly, leading to a reversal of $K_1/K_3$ from bend-dominated to splay-dominated elasticity.

The correlation lengths are anisotropic, with $\xi_Q^{\parallel}>\xi_Q^{\perp}$, while density--nematic coupling further renormalizes their magnitudes and anisotropy. As $\alpha$ increases and the system moves deeper into the nematic phase, the correlation length decreases. The parameter $u_0$ controls the density response: increasing $u_0$ suppresses density fluctuations and drives the system toward an effectively incompressible nematic dominated by nematic order fluctuations. Upon decreasing $u_0$, the coupled response develops a mean field nematic spinodal, where the correlation lengths diverge as $\xi\sim(u_0-u_{0,i})^{-1/2}$, with $u_{0,i}$ denoting the instability threshold.

SCFT calculations were also used to examine two-dimensional disclinations in a thin-film geometry. The anisotropic director profile is qualitatively consistent with the Frank elastic anisotropy obtained from linear response and with the classical Dzyaloshinskii solution. On the other hand, the effective elastic constant contrast depends on the distance to the defect core through its dependence on density and degree of order. The characteristic core size scales approximately linearly with the bulk correlation length $\xi_Q$ for localized defects. This relation is further supported by a density-coupled LdG model which we have introduced. In lyotropic chromonic liquid crystals, experimentally observed cores can be anomalously large relative to the mean aggregate length~\cite{zhou2017fine, re:zhou17b}, a regime not reached in the present model. The entropy reduction associated with chain conformation fluctuations in a nematic background that we have examined points to a possible resolution of this size disparity: a similar reduction may be mediated by molecular scission, a new microscopic degree of freedom that allows chain polydispersity.

Overall, the present work provides a microscopic framework linking molecular flexibility, density--nematic order coupling, elastic anisotropy, and defect structure in semiflexible nematics. The results demonstrate how collective continuum properties can emerge from microscopic chain statistics within a unified SCFT description. Several directions remain open for future work. In particular, the present results suggest possible extensions to the LdG theory to incorporate density variations, particularly in semiflexible and lyotropic nematic systems where density--nematic order coupling plays an important role. More generally, the framework developed here provides a basis for future extensions to self-assembled, polydisperse chains featuring reversible aggregation, as is the case in lyotropic chromonics.

\begin{acknowledgments}
This research has been supported by the National Science Foundation under contract DMR-2223707, by the Minnesota Supercomputing Institute of the University of Minnesota, and by the Advanced Cyberinfrastructure Coordination Ecosystem: Services \& Support (ACCESS) program, which is supported by U.S. National Science Foundation grants 2138259, 2138286, 2138307, 2137603, and 2138296.
\end{acknowledgments}


\appendix

\newcommand{\symrow}[2]{%
\noindent
\makebox[0.25\columnwidth][l]{#1}%
\parbox[t]{0.70\columnwidth}{\raggedright #2\par}%
\par\vspace{1pt}
}

\newcommand{\symheader}{%
\vspace{2pt}%
\noindent\hrule height 0.8pt
\vspace{1pt}%
\noindent\hrule height 0.4pt
\vspace{4pt}%
\noindent\makebox[0.25\columnwidth][l]{Symbol}%
\makebox[0.70\columnwidth][l]{Physical meaning}%
\par
\vspace{2pt}%
\noindent\hrule height 0.4pt
\vspace{4pt}%
}

\section{List of symbols}
\label{app:notation}
Only symbols used repeatedly throughout the manuscript are listed here; locally defined quantities are omitted.
\symheader

\symrow{$n$}{Number of wormlike chains.}
\symrow{$V$}{System volume.}
\symrow{$L$}{Chain contour length.}
\symrow{$l_p$}{Persistence length.}
\symrow{$s$}{Contour coordinate, $0\leq s\leq L$.}
\symrow{$\mathbf r$}{Spatial position.}
\symrow{$\mathbf u$}{Unit tangent vector along the chain contour.}
\symrow{$\hat{\rho}_u(\mathbf r,\mathbf u)$}{Microscopic orientation-resolved segment density.}
\symrow{$\rho_u(\mathbf r,\mathbf u)$}{Orientation-resolved segment-density field in the SCFT formulation.}
\symrow{$\rho(\mathbf r)$}{Scalar segment density obtained by orientational integration of $\rho_u(\mathbf r,\mathbf u)$.}
\symrow{$Q_{ij}(\mathbf r)$}{Density-weighted nematic order tensor.}
\symrow{$S(\mathbf r)$}{Scalar uniaxial nematic order parameter.}
\symrow{$P(\mathbf r)$}{Biaxial order parameter.}
\symrow{$\mathbf n(\mathbf r)$}{Local nematic director.}
\symrow{$u_0$}{Strength of the orientation-independent segment--segment repulsion.}
\symrow{$u_2$}{Strength of the Maier--Saupe orientational interaction.}
\symrow{$w(\mathbf r,\mathbf u)$}{Auxiliary field conjugate to $\rho_u(\mathbf r,\mathbf u)$.}
\symrow{$q(\mathbf r,\mathbf u,s)$}{Single-chain propagator satisfying the modified diffusion equation.}
\symrow{$Z_1[w]$}{Single-chain partition function in the auxiliary field $w$.}
\symrow{$Z_Q(w_Q)$}{Orientational part of the single-chain partition function after factoring out the isotropic contribution.}
\symrow{$g(S;L/l_p)$}{Conformational free-energy contribution per chain associated with nematic ordering, relative to the isotropic state.}

\symrow{$\alpha$}{Effective alignment strength, $\alpha \equiv u_2\rho L=nL^2u_2/V$.}
\symrow{$f_{\rm eff}(\rho,S)$}{Effective free-energy density for a spatially uniform uniaxial state.}
\symrow{$\mu$}{Chain chemical potential.}
\symrow{$\Pi$}{Osmotic pressure.}
\symrow{$\Omega$}{Grand potential.}
\symrow{$\omega$}{Grand-potential density, $\omega=\Omega/V$.}
\symrow{$S_N$}{Equilibrium scalar order parameter of the nematic state.}
\symrow{$\alpha_I,\alpha_N$}{Effective alignment strengths of the isotropic and nematic phases at coexistence.}
\symrow{$\alpha_{I,\mathrm{sp}},\alpha_{N,\mathrm{sp}}$}{Isotropic and nematic spinodal values of $\alpha$.}
\symrow{$\mathbf k$}{Wavevector of a spatial perturbation, with $k=|\mathbf k|$.}
\symrow{$h_\rho,h_Q$}{Strength of external fields coupled to density and density-weighted nematic order, respectively.}
\symrow{$Q(\mathbf r)$}{Scalar projection of the nematic order tensor used in the coupled response; for an aligned uniaxial state, $Q=2\rho S/3$.}
\symrow{$\boldsymbol{\chi}(\mathbf k)$}{Susceptibility matrix for coupled density--nematic-order fluctuations.}
\symrow{$\mathbf M(\mathbf k)$}{Inverse susceptibility matrix, $\mathbf M(\mathbf k)=\boldsymbol{\chi}^{-1}(\mathbf k)$.}
\symrow{$A_{ij}$}{Zero-wavevector coefficients of the inverse susceptibility or, in the uniform theory, elements of the free-energy Hessian; $i,j$ label $(\rho,Q)$ in the linear-response analysis and $(\rho,S)$ in the uniform theory.}
\symrow{$K_{ij}^{\alpha\beta}$}{Gradient coefficients in the long-wavelength expansion of $M_{ij}(\mathbf k)$; $i,j$ label $(\rho,Q)$ and $\alpha,\beta$ denote spatial coordinates.}
\symrow{$K_{ij}^{\parallel},K_{ij}^{\perp}$}{Gradient coefficients for wavevectors parallel and perpendicular to the nematic director, respectively.}
\symrow{$\lambda_\pm$}{Eigenvalues of the coupled inverse susceptibility matrix; $\lambda_-$ denotes the soft mode.}
\symrow{$A_{Q,-}$}{Zero-wavevector coefficients of the projected nematic-order response ($Q$) and the soft mode ($-$), respectively.}
\symrow{$K_{Q,-}^{\parallel,\perp}$}{Gradient coefficients of the projected nematic-order response ($Q$) and the soft coupled mode ($-$), for variations parallel and perpendicular to the nematic director, respectively.}
\symrow{$\xi_{Q,-}^{\parallel,\perp}$}{Correlation lengths associated with the projected nematic-order response ($Q$) and the soft mode ($-$), for variations parallel and perpendicular to the nematic director, respectively.}
\symrow{$K_1,K_2,K_3$}{Frank elastic constants for splay, twist, and bend, respectively.}
\symrow{$\chi_i(k)$}{Director susceptibility for Frank mode $i=1,2,3$.}
\symrow{$r_\parallel,r_\perp$}{Defect-core radii associated with gradients parallel and perpendicular to the local director.}
\symrow{$\epsilon$}{Frank elastic anisotropy, $\epsilon=(K_3-K_1)/(K_1+K_3)$.}

\vspace{2pt}
\noindent\hrule height 0.4pt
\vspace{1pt}
\noindent\hrule height 0.8pt

\section{Grand potential formulation}
\label{app:binodals and spinodals}
The grand potential density is defined as
\begin{equation}
\beta\omega(\mu,S)=
\beta f_{\rm eff}(\rho,S)-
\frac{\beta\mu}{L}\rho ,
\end{equation}
with the density $\rho$ determined implicitly by $ \partial(\beta f_{\rm eff})/\partial\rho= \beta\mu/L.$ The osmotic pressure $\beta \Pi=\rho \partial(\beta f_{\rm eff})/\partial\rho-\beta f_{\rm eff}$ implies $\beta\omega=-\beta \Pi.$ Therefore, equality of osmotic pressure is equivalent to equality of the grand potential density, establishing the equivalence between the coexistence conditions.

At fixed $\mu$, the curvature with respect to $S$ is
\begin{equation}
\left(
\frac{\partial^2 \beta\omega}{\partial S^2}
\right)_{\mu}=A_{SS}-
\frac{A_{\rho S}^2}{A_{\rho\rho}},
\end{equation}
where $A_{ij}$ are the Hessian matrix elements defined in the main text. Since $A_{\rho\rho}>0$ throughout the physically relevant parameter range, the spinodal condition $\partial^2(\beta\omega)/\partial S^2=0$ is equivalent to $\det A=0.$ Thus, the spinodal obtained from the vanishing curvature of the grand potential density at fixed chemical potential is identical to the spinodal determined from the Hessian stability criterion.

\section{Numerical method for SCFT}
\label{appendix:numerics}

The SCFT equations, Eqs.~(\ref{eq:sc_rho}) and (\ref{eq:sc_w}), are solved numerically using a real spherical-harmonic expansion together with an iterative procedure, following the numerical methods developed in our previous work~\cite{qing2025self} and in Ref.~\cite{spencer2020nematic}. The Python code used in this work is available at Zenodo~\cite{qing2026scft}.

\subsection{Real spherical harmonic representation and iteration method}
An orientation dependent field is expanded in real spherical harmonics as
\begin{equation}
f(\mathbf r,\mathbf u)
=
\sum_{l,m}
f_l^m(\mathbf r)\,
\tilde{Y}_l^m(\mathbf u),
\end{equation}
where $f$ represents fields such as the propagator $q(\mathbf r,\mathbf u,s)$, orientation-resolved density field $\rho_u(\mathbf r,\mathbf u)$, and auxiliary field $w(\mathbf r,\mathbf u)$, with the expansion for the propagator understood to be performed at fixed contour variable $s$. In the numerical implementation, we generally truncate the spherical-harmonic expansion at $l=4$ and discretize the spatial domain on a Cartesian grid. For the Frank-elasticity results shown in Fig.~\ref{fig:Ki_alpha}, we use the higher truncation $l=6$ to ensure improved numerical accuracy. For linear response calculations, spatial variations are restricted to one dimension, such that all fields depend only on $x$, i.e., $f_l^m=f_l^m(x_i)$, with the $y$ and $z$ directions assumed to be uniform. For defect calculations, spatial variations are resolved in two dimensions, and fields are represented on a grid $(x_i,y_j)$, while the $z$ direction remains uniform, i.e., $f_l^m=f_l^m(x_i,y_j)$. Although spatial variations are restricted to one or two dimensions, the orientational degrees of freedom remain fully three-dimensional through the real spherical-harmonic expansion. The corresponding tensor order parameter $\mathbf Q$ is therefore defined in three dimensions, although certain components may vanish due to the symmetry of the imposed geometry.

The contour variable is defined over the interval $s\in[0,L]$, and the propagator $q_l^m(\mathbf r,s)$ is obtained by solving the MDE using a Lax--Wendroff scheme~\cite{daoulas2005self,qing2025self} with a uniform contour step size $\Delta s=L/200$. Picard iteration with simple mixing~\cite{jiang2010isotropic,deng2010wormlike} is employed, with the fields updated according to
\begin{equation}
w^{(i+1)}=(1-\gamma)w^{(i)}+\gamma w_{\mathrm{new}}^{(i)},
\end{equation}
where $i$ labels the iteration step and $\gamma$ is the mixing parameter. Stable convergence typically requires a relatively small value of $\gamma$ (normally $\gamma<0.1$), which leads to slow convergence. To improve convergence efficiency while maintaining stability, a periodically enhanced mixing strategy is adopted, in which a small baseline value of $\gamma$ is used for most iterations, while a larger value is temporarily applied for several iterations before reverting to the baseline value.

\subsection{Linear response calculations}
For linear response calculations, the external fields are imposed with a sinusoidal modulation along the $x$ direction. The computations are performed for wavelengths $\lambda \sim 10$--$40L$, corresponding to the small-$k$ regime. In the numerical implementation, the system size $L_x$ is taken to be equal to the imposed wavelength, and periodic boundary conditions are applied. The domain in the $x$ direction is discretized into 200 uniformly spaced grid points. The amplitude of the prescribed external fields is chosen in the range $h \sim 10^{-6}$--$10^{-5}$. We have verified that, within this range, the induced response amplitude scales linearly with the applied field for fixed $k$, ensuring that all calculations are performed within the linear response regime. 

In the linear response calculations, convergence can be further accelerated by temporarily adjusting the amplitude of external field during the iteration: when the instantaneous response amplitude remains below the target value, a larger field amplitude is applied to drive the response upward more rapidly, while a smaller field amplitude is used when the response exceeds the target value. The adjustment range is progressively reduced, and the external field is repeatedly restored to the prescribed value until the response converges to the desired accuracy.
\subsection{Defect calculations}
For the topological defect calculations, the system is defined in a two-dimensional square domain with lateral dimensions $L_x = L_y \sim 4$--$10L$, discretized into $120 \times 120$ uniform grid points. Homogeneous Neumann boundary conditions are imposed at the domain boundaries. We adopt a similar initialization procedure as in Ref.~\cite{qing2025self}, specifying the initial configuration by a radially varying scalar order parameter and a prescribed director field corresponding to a $+1/2$ disclination. The initial director fields are constructed using the values of $\epsilon$ extracted from the linear response analysis. The system is then relaxed to equilibrium through the iterative procedure. We also tested initial director fields with $\epsilon=0$ and observed the development of director anisotropy during iteration, with the same qualitative angular dependence.

\section{Validation and directional dependence of the coupled linear response}
\label{app:lr_details}

\subsection{Isotropic reference state}

In the isotropic phase, no preferred director exists in the absence of external fields. A spatially varying density field $h_\rho(\mathbf r)$ nevertheless induces a weak nematic response whose symmetry axis is selected by the wavevector $\mathbf k$. We therefore choose the easy axis of the field coupled to $Q$ parallel to $\mathbf k$.

\begin{figure}[t]
\centering
\includegraphics[width=1\columnwidth]{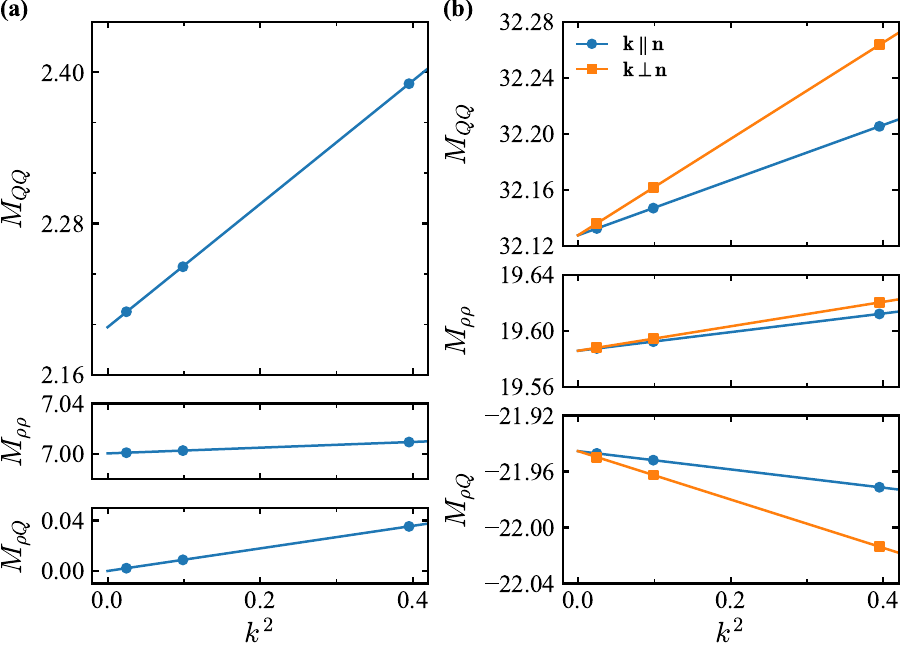}
\caption{Elements of the inverse susceptibility matrix $M_{ij}$ as functions of $k^2$ for representative (a) isotropic and (b) nematic reference states. The parameters are $L=l_p=1$, $n/V=1$, $u_0=6$, and $u_2=10$ in (a), and $L=l_p=1$, $n/V=1.3$, $u_0=10$, and $u_2=10$ in (b). Symbols denote numerical results and solid lines denote linear fits to the small-$k$ form in Eq.~\eqref{eq:M_small_k_app}.}
\label{fig:M_elements_q2_app}
\end{figure}

Figure~\ref{fig:M_elements_q2_app}(a) shows a representative small-$k$ fit of the inverse susceptibility matrix $\mathbf M(\mathbf k)=\boldsymbol{\chi}^{-1}(\mathbf k)$. In particular, the off-diagonal element extrapolates to $M_{\rho Q}(k\to0)=0$ within numerical accuracy, consistent with $A_{\rho Q}=0$ at $S=0$ in Eq.~\eqref{eq:A_ij_theory}. The gradient coupling remains finite, however, with $K_{\rho Q}>0$. Thus, while uniform density and nematic-order variations are decoupled in the isotropic state, spatially varying perturbations generate a finite density--order coupling at order $k^2$.

\begin{table} [h]
\centering
\caption{Linear response coefficients in the isotropic reference state under different parameter values. Reference parameters: $L=l_p=1$, $n/V=1$, $u_0=6$, $u_2=10$.}
\label{tab:iso_response}
\begin{tabular}{lccccccc}
\toprule
Case & $\alpha$ & $A_{\rho\rho}$ & $K_{\rho\rho}$ & $A_{\rho Q}\times10^{5}$ & $K_{\rho Q}$ & $A_{QQ}$ & $K_{QQ}$ \\
\midrule
Reference      & 10  & 7.00 & 0.0228 & $-3.94$ & 0.0896 & 2.197 & 0.4902 \\
$n/V=0.5$    & 5   & 8.00 & 0.0456 & $-7.86$ & 0.1793 & 26.90 & 0.9803 \\
$n/V=0.25$   & 2.5 & 10.0 & 0.0912 & $-16.6$ & 0.3586 & 76.30 & 1.9605 \\
$u_2=5$      & 5   & 7.00 & 0.0228 & $-3.93$ & 0.0896 & 13.45 & 0.4901 \\
$L=l_p=0.5$  & 2.5 & 10.0 & 0.0224 & $-0.797$ & 0.0879 & 76.29 & 0.4818 \\
$u_0=5.7$    & 10  & 6.70 & 0.0228 & $-3.38$ & 0.0896 & 2.197 & 0.4902 \\
\bottomrule
\end{tabular}
\end{table}

The numerical values of the extracted coefficients under several parameter variations are summarized in Table~\ref{tab:iso_response}. In the isotropic state, $S=0$, Eq.~\eqref{eq:A_ij_theory} reduces to
\begin{equation}
\begin{aligned}
A_{\rho\rho} &= u_0+\left(\frac{nL^2}{V}\right)^{-1},
\qquad
A_{\rho Q}=0,\\
A_{QQ} &= -\frac{9}{4}u_2
+\frac{9}{4}\left(\frac{nL^2}{V}\right)^{-1}g''(0;L/l_p).
\end{aligned}
\label{eq:A_iso_app}
\end{equation}
The numerical results are consistent with these expressions: $A_{\rho Q}$ remains negligible within numerical accuracy, while the dependence of $A_{\rho\rho}$ and $A_{QQ}$ on $u_0$, $u_2$, $n/V$, and $L$ follows Eq.~\eqref{eq:A_iso_app}. The gradient coefficients $K_{ij}$ are essentially insensitive to $u_0$ and $u_2$ and change little when $L$ is varied at fixed $L/l_p$, whereas reducing the chain density $n/V$ increases all $K_{ij}$ approximately in proportion to $(n/V)^{-1}$.

\subsection{Directional dependence in the nematic state}

In the nematic phase, the uniform state possesses a spontaneously selected director $\mathbf n$, and the response depends on the orientation of the wavevector relative to $\mathbf n$. We therefore consider the two representative geometries $\mathbf k\parallel\mathbf n$ and $\mathbf k\perp\mathbf n$. The corresponding inverse susceptibility is described by
\begin{equation}
M_{ij}(\mathbf k)=A_{ij}+K_{ij}^{\parallel}k_{\parallel}^2+K_{ij}^{\perp}k_{\perp}^2+O(k^4),
\label{eq:M_small_k_app}
\end{equation}
which follows from the small-$k$ expansion in Eq.~\eqref{eq:M_small_k} together with the uniaxial decomposition of $K_{ij}^{\alpha\beta}$ in Eq.~\eqref{eq:K_perp_para}.

Figure~\ref{fig:M_elements_q2_app}(b) shows representative fits for the two directions. The numerical data follow the expected quadratic dependence on wavevector. The fitted intercepts $A_{ij}$ are independent of the direction of $\mathbf k$ within numerical accuracy. The gradient coefficients, by contrast, exhibit clear anisotropy, with $K_{ij}^{\parallel}\neq K_{ij}^{\perp}$. The off-diagonal elements are comparable in magnitude to the diagonal ones, demonstrating that density and nematic order remain strongly coupled in the nematic phase. Moreover, $A_{\rho Q}$ and $K_{\rho Q}$ are both negative, implying that the system energetically favors in-phase variations of $\rho$ and $Q$.

An important consequence is that the anisotropy of the correlation lengths cannot be inferred from the diagonal gradient coefficients alone. Although the diagonal coefficients satisfy $K_{\rho\rho}^{\parallel}<K_{\rho\rho}^{\perp}$ and $K_{QQ}^{\parallel}<K_{QQ}^{\perp}$ over the parameter range considered here, the full coupled response yields $\xi_Q^{\parallel}>\xi_Q^{\perp}$. This reversal arises from the off-diagonal density--order coupling, which mixes density and nematic order fluctuations and changes the effective stiffness of the collective mode.

\section{Density-coupled Landau model for defect core scaling}
\label{app:ldg_core}

To examine whether the approximate linear relation between defect core size and equilibrium correlation length found in the SCFT calculations is specific to the microscopic wormlike chain model, we consider a minimal density-coupled LdG model. The nematic order is restricted to a locally uniaxial form, so biaxial reconstruction near the defect center is not resolved, and the elastic free energy is taken in the one-constant approximation, neglecting elastic anisotropy. Nevertheless, the model retains the scalar order recovery, density redistribution, and their gradient penalties that control the characteristic radial extent of the core, making it sufficient for testing the relation between core radius and correlation length.

\subsection{Density-coupled Landau free energy}
\label{app:ldg_model}

The local free energy density is motivated by the structure of the SCFT effective free energy in Eq.~\eqref{eq:eff_f}. Rather than retaining a quantitative mapping to the microscopic SCFT parameters, we use a dimensionless phenomenological model in which the factors involving the chain length $L$ and the overall free energy scale are absorbed into the coefficients defined below. The $S$-dependent contribution $g(S;L/l_p)$ is replaced by a Landau free energy,
\begin{equation}
g_{\rm LdG}(S)
=
\frac{a}{2}S^2
-\frac{b}{3}S^3
+\frac{c}{4}S^4 ,
\label{eq:ldg_g}
\end{equation}
with $b>0$ and $c>0$. The local free energy density is then
\begin{equation}
f_{\rm LdG}(\rho,S)
=
\frac{\kappa_\rho}{2}\rho^2
+
\rho g_{\rm LdG}(S)
-
\frac{u_2}{2}\rho^2S^2 .
\label{eq:ldg_flocal}
\end{equation}
The quadratic density term provides a minimal local approximation to the density dependent part of the SCFT free energy. Expanding the SCFT density contribution about a reference state gives an effective density stiffness, which is represented phenomenologically here by $\kappa_\rho$.

Spatial variations are described by
\begin{equation}
f_{\rm grad}
=
\frac{K_\rho}{2}|\nabla\rho|^2
+
\frac{K_S}{2}|\nabla S|^2
+
\frac{3K_S}{2}S^2
(\partial_k n_i)(\partial_k n_i).
\label{eq:ldg_fgrad}
\end{equation}
The last two terms correspond to the usual one-constant LdG gradient free energy after restricting the nematic order tensor to a locally uniaxial form. For simplicity, we neglect mixed density--nematic order gradient couplings, which are present in the full SCFT response.

\subsection{Bulk correlation length}
\label{app:ldg_xi}

We characterize small fluctuations about a uniform nematic state using the same coupled response construction introduced in Sec.~\ref{subsec:lr_coupling}. The only difference is the choice of nematic variable: whereas the SCFT response is formulated in terms of the density-weighted quantity $Q$, the natural variables of the present LdG model are $(\rho,S)$.

The inverse susceptibility matrix is obtained from the local Hessian and gradient terms of Eqs.~\eqref{eq:ldg_flocal} and \eqref{eq:ldg_fgrad}, with the long-wavelength expansion constructed
in the same manner as Eq.~\eqref{eq:M_small_k}. Minimizing the free energy with respect to the density fluctuation yields an effective response for the $S$ mode and defines the associated correlation length $\xi_S$, in direct analogy with the projected correlation length $\xi_Q$ defined in Sec.~\ref{subsec:lr_coupling}.

\subsection{Radial scalar profiles for a $+1/2$ defect}
\label{app:ldg_defect}

We next apply the density-coupled Landau model to a $+1/2$ disclination. We assume radially symmetric density and scalar order profiles, together with the director configuration under one-constant approximation,
\begin{equation}
\mathbf n=\left(\cos\frac{\varphi}{2},\sin\frac{\varphi}{2},0\right),
\qquad
\rho=\rho(r),
\qquad
S=S(r).
\end{equation}
Here $(r,\varphi)$ are polar coordinates centered at the defect core, consistent with the notation used in the main text. For this director configuration, $(\partial_k n_i)(\partial_k n_i)=1/(4r^2)$. The calculation is performed in a cylindrical cell of radius $R$ at fixed mean density,
\begin{equation}
\bar\rho=\frac{2}{R^2}\int_0^R r\,dr\,\rho(r).
\label{eq:ldg_mass}
\end{equation}
Introducing the chemical potential $\mu$ as a Lagrange multiplier to impose the mass constraint in Eq.~\eqref{eq:ldg_mass}, stationarity of the constrained free energy functional with respect to $\rho(r)$ and $S(r)$ yields the Euler--Lagrange equations
\begin{equation}
-K_\rho\left(\rho''+\frac{\rho'}{r}\right)+\kappa_\rho\rho+g_{\rm LdG}(S)-u_2\rho S^2-\mu=0,
\label{eq:ldg_el_rho}
\end{equation}
\begin{equation}
-K_S\left(S''+\frac{S'}{r}\right)+\rho g_{\rm LdG}'(S)-u_2\rho^2S+\frac{3K_S}{4r^2}S=0.
\label{eq:ldg_el_S}
\end{equation}
Here $g_{\rm LdG}'$ denotes differentiation with respect to $S$, while primes on $\rho$ and $S$ denote derivatives with respect to $r$. At the defect center, regularity of the density and suppression of nematic order require
\begin{equation}
\rho'(0)=0,
\qquad
S(0)=0.
\label{eq:ldg_bc_center}
\end{equation}
At the outer boundary we impose the conditions
\begin{equation}
\rho'(R)=0,
\qquad
S'(R)=0.
\label{eq:ldg_bc_outer}
\end{equation}
The profiles $\rho(r)$ and $S(r)$ and the Lagrange multiplier $\mu$ are determined simultaneously by Eqs.~\eqref{eq:ldg_mass}--\eqref{eq:ldg_bc_outer}, with the boundary values $\rho(R)$ and $S(R)$ emerging as part of the solution rather than being imposed independently.

For comparison with the correlation length $\xi_S$, we define the defect-core radius from the recovery of the scalar order parameter,
\begin{equation}
S(r_{\rm core})=0.95\,S(R).
\label{eq:ldg_rcore}
\end{equation}
The correlation length $\xi_S$ associated with each defect solution is evaluated at the corresponding bulk nematic state represented by $(\rho(R),S(R))$.
\begin{figure*}[t]
\centering
\includegraphics[width=0.95\textwidth]{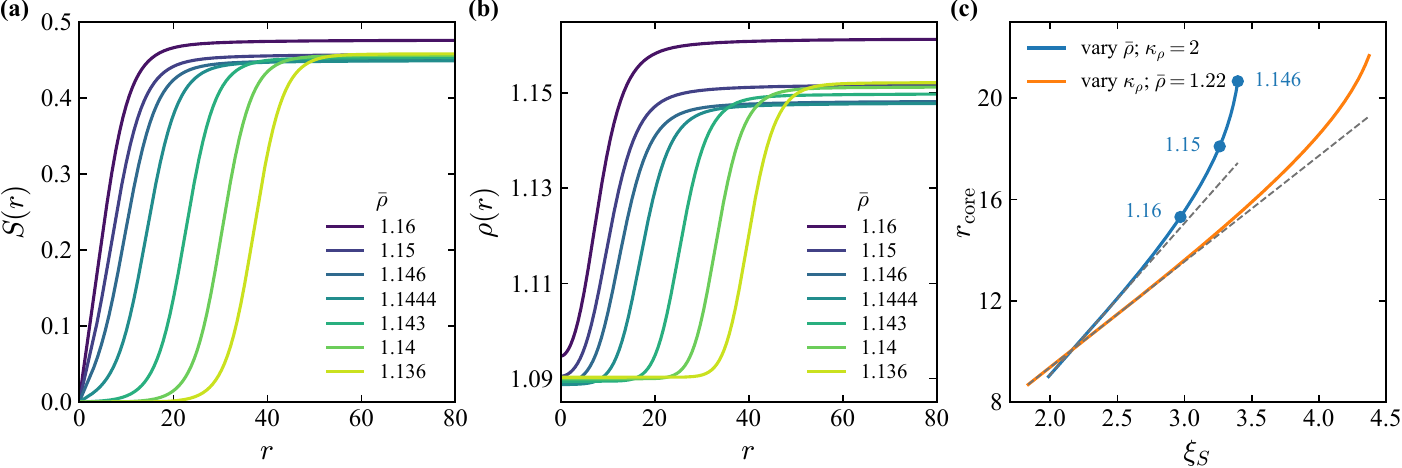}
\caption{
Density-coupled LdG model calculation of defect-core structure and its relation to the bulk correlation length. (a) Uniaxial order profiles $S(r)$ and (b) density profiles $\rho(r)$ for decreasing mean density $\bar\rho$ at fixed $\kappa_\rho=2$. (c) Defect core radius $r_{\rm core}$ as a function of the bulk correlation length $\xi_S$, evaluated at the corresponding outer nematic state. The two curves correspond to varying $\bar\rho$ at fixed $\kappa_\rho=2$ and varying $\kappa_\rho$ at fixed $\bar\rho=1.22$. The labeled points at $\bar\rho=1.16$, $1.15$, and $1.146$ correspond to representative profiles shown in panels (a) and (b). The dashed lines indicate the linear fit in the well localized core regime. Deviations from this behavior become more pronounced as the low-order region broadens toward the crossover. Other parameters are $a=1.2$, $b=0.7$, $c=1.3$, $u_2=K_\rho=K_S=1$, and $R=80$.
}
\label{fig:ldg_core_scaling}
\end{figure*}
\subsection{Localized-core scaling and approach to crossover}
\label{app:ldg_scaling}
The numerical results below use $a=1.2$, $b=0.7$, $c=1.3$, $u_2=K_\rho=K_S=1$, and $R=80$. Figure~\ref{fig:ldg_core_scaling}(a,b) first illustrates the evolution of the defect structure upon decreasing the mean density $\bar\rho$ at fixed $\kappa_\rho=2$. For the larger values of $\bar\rho$, the suppression of both $S(r)$ and $\rho(r)$ remains confined to a localized region near the defect center. For the present parameter set, as $\bar\rho$ is decreased, this low-order region broadens continuously and eventually develops into an extended isotropic domain, or tactoid-like, structure. We refer to this smooth evolution as a localized-to-domain-like crossover.

The relation between the defect core size and the bulk correlation length is examined in Fig.~\ref{fig:ldg_core_scaling}(c) along two parameter paths: varying $\bar\rho$ at fixed $\kappa_\rho=2$, and varying $\kappa_\rho$ at fixed $\bar\rho=1.22$. For well localized defects, $r_{\rm core}$ exhibits an approximately linear dependence on $\xi_S$ along both paths, although the numerical slope depends on the parameter path. As $\bar\rho$ is decreased and the low-order region progressively broadens, the data begin to deviate upward from the linear behavior observed for the more localized cores.

The common feature of the two parameter scans is therefore not a universal proportionality factor, but an approximately linear scaling of $r_{\rm core}$ with $\xi_S$ within the well localized core regime.


\bibliography{reference}

\end{document}